\documentclass[11pt]{article}
\usepackage[margin=1in]{geometry}
\usepackage{times}

\usepackage{xcolor}
\usepackage[colorlinks=true,citecolor=blue,linkcolor=blue]{hyperref}

\hypersetup{
    colorlinks,
    linkcolor={red!50!black},
    citecolor={blue!50!black},
    urlcolor={blue!80!black}
}

\usepackage[]{amsmath,amssymb,amsfonts,latexsym,amsthm,enumerate,fullpage,xcolor,bbm,breakcites}
\usepackage[nameinlink, noabbrev, capitalize]{cleveref}
\usepackage{comment}
\usepackage{nicefrac}
\usepackage{thm-restate}
\crefname{prop}{Proposition}{Propositions}
\crefname{claim}{Claim}{Claims}
\crefname{ineq}{inequality}{inequalities}
\creflabelformat{ineq}{#2(#1)#3}

\newtheorem{theorem}{Theorem}
\newtheorem{lemma}{Lemma}

\newtheorem{claim}{Claim}
\newtheorem{fact}{Fact}

\newtheorem{corollary}{Corollary}
\newtheorem{definition}{Definition}

\crefname{THM}{Theorem}{Theorems}

\usepackage{subcaption}
\usepackage{graphicx}
\usepackage{tikz}
\usepackage{color, colortbl}
\definecolor{LightCyan}{rgb}{0.88,1,1}
\definecolor{Gray}{gray}{0.9}
\usetikzlibrary{positioning}
\usetikzlibrary{decorations.pathmorphing}
\usetikzlibrary{automata,positioning}
\usetikzlibrary{arrows}
\usetikzlibrary{arrows.meta}
\usetikzlibrary{calc}

\newcommand{\E}{\mathbb{E}}
\newcommand{\N}{\mathbb{N}}

\newcommand{\F}{\mathbb{F}}

\newcommand{\supp}{\mathsf{supp}}
\newcommand{\pr}{{\prime}}
\newcommand{\prpr}{{\prime\prime}}
\newcommand{\U}{\mathbf{U}}
\newcommand{\X}{\mathbf{X}}
\newcommand{\Y}{\mathbf{Y}}

\newcommand{\A}{\mathbf{A}}
\newcommand{\B}{\mathbf{B}}

\newcommand{\W}{\mathbf{W}}

\newcommand{\ZZ}{\mathbf{Z}}

\newcommand{\zo}{\{0,1\}}

\newcommand{\eps}{\varepsilon}

\newcommand{\Ext}{\mathsf{Ext}}
\newcommand{\Disp}{\mathsf{Disp}}

\newcommand{\eval}{\operatorname{eval}}
\newcommand{\rank}{\operatorname{rank}}
\newcommand{\spanvec}{\operatorname{span}}

\newcommand{\cC}{\mathcal{C}}
\newcommand{\cB}{\mathcal{B}}

\newcommand{\minH}{H_\infty}

\DeclareMathOperator{\bias}{bias}

\usepackage{subfiles}

\newcommand{\dobib}{
    \bibliographystyle{alpha}
    \bibliography{references} 
}

\begin{document}
\renewcommand{\dobib}{}

\title{Low-Degree Polynomials Are Good Extractors}

\author{Omar Alrabiah\thanks{UC Berkeley. \texttt{oalrabiah@berkeley.edu}. Supported by a Saudi Arabian Cultural Mission (SACM) Scholarship, NSF Award CCF-2210823, and a Simons Investigator Award (Venkatesan Guruswami).} \and Jesse Goodman\thanks{UT Austin. \texttt{jpmgoodman@utexas.edu}. Supported by a Simons Investigator Award (\#409864, David Zuckerman).} \and Jonathan Mosheiff\thanks{Ben-Gurion University. \texttt{mosheiff@bgu.ac.il}. Supported by an Alon Fellowship and by DOE grant \# DE-SC0024124 while visiting the Simons Institute for the Theory of Computing}\and Jo\~ao Ribeiro\thanks{Instituto de Telecomunicações and Departamento de Matemática, Instituto Superior Técnico, Universidade de Lisboa. \texttt{jribeiro@tecnico.ulisboa.pt}. Work mainly done while at NOVA LINCS and NOVA School of Science and Technology, and while visiting the Simons Institute for the Theory of Computing supported by DOE grant \# DE-SC0024124. Also supported by NOVA LINCS (ref.\ UIDB/04516/2020) and by
FCT/MECI through national funds and when applicable co-funded EU funds under UID/50008: Instituto de Telecomunicações.}}

\date{}
\maketitle

\begin{abstract}

We prove that random low-degree polynomials (over $\mathbb{F}_2$) are unbiased, in an extremely general sense. That is, we show that random low-degree polynomials are good \emph{randomness extractors} for a wide class of distributions. Prior to our work, such results were only known for the small families of (1) uniform sources, (2) affine sources, and (3) local sources. We significantly generalize these results, and prove the following.

\begin{enumerate}
    \item \textbf{Low-degree polynomials extract from small families.} We show that a random low-degree polynomial is a good low-error extractor for \emph{any} small family of sources. In particular, we improve the positive result of Alrabiah, Chattopadhyay, Goodman, Li, and Ribeiro (ICALP 2022) for local sources, and give new results for polynomial and variety sources via a single unified approach.
    
    \item \textbf{Low-degree polynomials extract from sumset sources.} We show that a random low-degree polynomial is a good extractor for sumset sources, which are the most general \emph{large} family of sources (capturing independent sources, interleaved sources, small-space sources, and more). Formally, for any even \(d\), we show that a random degree \(d\) polynomial is an \(\eps\)-error extractor for \(n\)-bit sumset sources with min-entropy \(k=O(d(n/\eps^2)^{2/d})\). This is nearly tight in the polynomial error regime.
\end{enumerate}

Our results on sumset extractors imply new complexity separations for linear ROBPs, and the tools that go into its proof may be of independent interest. The two main tools we use are a new structural result on sumset-punctured Reed-Muller codes, paired with a novel type of reduction between extractors. Using the new structural result, we obtain new limits on the power of sumset extractors, strengthening and generalizing the impossibility results of Chattopadhyay, Goodman, and Gurumukhani (ITCS 2024).

\dobib

\end{abstract}

\newpage

\tableofcontents

\newpage

\section{Introduction}

In this work, we are interested in the following open-ended question:
\begin{center}
    \emph{How biased is a random degree \(d\) polynomial \(f:\F_2^n\to \F_2\)?}
\end{center}
By \emph{random} degree \(d\) polynomial, we mean a polynomial of the form
\[
    f(x)=\sum_{S\subseteq[n]:|S|\leq d}\alpha_S\cdot x^S,
\]
where \(x^S:=\prod_{i\in S}x_i\) and the coefficients \(\alpha_S\) are sampled independently and uniformly at random from \(\F_2\). And by \emph{bias}, perhaps the most basic definition is to simply check the difference between how many inputs are mapped to \(0\) and \(1\). Or more formally, letting \(\U_n\) denote the uniform distribution over \(\F_2^n\),
\[
\bias(f):=\Pr_{x\sim\U_n}[f(x)=0]-\Pr_{x\sim\U_n}[f(x)=1].
\]
For these definitions of \emph{random} and \emph{bias}, Ben-Eliezer, Hod, and Lovett \cite{BHL12} -- building on a pair of earlier papers \cite{KS05,ABK08} -- provided a complete answer to the above question. In particular, they showed sharp concentration bounds on \(|\bias(f)|\), concluding that a random degree \(d\) polynomial is essentially unbiased on a uniform input, with extremely high probability. More precisely, they proved the following.

\setcounter{theorem}{-1}

\begin{theorem}[Random low-degree polynomials are unbiased \protect{\cite[Lemma 1.2]{BHL12}}]\label{intro:thm:BHL12}
For every \(\delta\in(0,1)\) there is a constant \(c>0\) such that the following holds. Let \(d\in\N\) be an integer satisfying \(1\leq d\leq(1-\delta)k\). Then for a random degree \(d\) polynomial \(f:\F_2^n\to \F_2\),
\[
\Pr_f\left[|\bias(f)|>2^{-cn/d}\right]\leq2^{-c\binom{n}{\leq d}}.
\]
\end{theorem}
\noindent They showed these bounds were tight, and a later work extended these results to all prime fields \cite{BOY20}.

While \cref{intro:thm:BHL12} is interesting in its own right (since low-degree polynomials are fundamental objects), its pursuit was largely motivated by applications in coding theory, complexity theory, and pseudorandomness. Indeed, given this new result, Ben-Eliezer, Hod, and Lovett immediately obtained important corollaries in each of these areas. In coding theory, they obtained new tail bounds on the weight distribution of Reed-Muller codes. In complexity theory, they showed that (random) low-degree polynomials cannot be approximated by polynomials of smaller degree. And in pseudorandomness, they obtained new lower bounds on the seed length of pseudorandom generators (PRGs) for low-degree polynomials.

Since we now understand the bias of a random low-degree polynomial on a \emph{uniform} input, it is natural to ask whether a more general result can be proven for \emph{weakly random} inputs - especially given the connection this problem has to pseudorandomness. But in order to make this question formal, we'll need a more general definition of \emph{bias}, which allows the function to receive a weakly random input. Towards this end, given a random variable (``source'') \(\X\) over \(\F_2^n\), we define
\[
    \bias_\X(f):=\Pr_{x\sim\X}[f(x)=0]-\Pr_{x\sim\X}[f(x)=1].
\]
Given this definition, a simple observation is that \(\bias_{\U_n}(f)=\bias(f)\), and thus establishing concentration bounds for \(|\bias_\X(f)|\) is a strictly more general problem than doing so for \(|\bias(f)|\). But how should we use this generality? And for what distributions \(\X\) is it actually interesting to understand \(|\bias_\X(f)|\)? Recall that we would like to understand \(|\bias_\X(f)|\) for \emph{weakly random} \(\X\), but it is still not clear what weakly random should mean. To answer all of these questions and more, we enter the world of \emph{randomness extraction}.

\subsubsection*{Randomness extractors}

Randomness extractors are fundamental objects in pseudorandomness and complexity theory. They are motivated by the fact that nature can only provide us with weak sources of randomness, yet most applications in computer science require perfectly uniform bits. Formally, they are defined as follows.

\begin{definition}[Randomness extractor]\label{def:extractor}
    Let \(\mathcal{X}\) be a family of sources \(\X\sim\zo^n\). A deterministic function \(\Ext:\zo^n\to\zo^m\) is an \emph{extractor} for \(\mathcal{X}\) with error \(\eps\) if for any \(\X\in\mathcal{X}\),
    \begin{align*}\label{eq:extractor-condition}
    \Delta(\Ext(\X),\U_m)\leq\eps,
    \end{align*}
    where $\Delta(\cdot,\cdot)$ denotes statistical distance. For short, we also call \(\Ext\) an \(\eps\)-extractor for \(\mathcal{X}\).
\end{definition}
\noindent Ever since extractors were first introduced, they have found countless unexpected applications in complexity, cryptography, pseudorandomness, and theoretical computer science. For a survey, see \cite{Vad12,Goo23}.

In this paper, we will focus on extractors that output one bit.\footnote{As we will see, there are often relatively standard tricks that can boost the output length of an extractor, once one bit is obtained.} In this case, the requirement in \cref{def:extractor} reduces to a requirement that for any \(\X\in\mathcal{X}\), it holds that \(|\Pr[\Ext(\X)=1]-1/2|\leq\eps\). Or in other words,
\[
|\bias_\X(\Ext)|\leq2\eps.
\]
Thus, returning to our original discussion, we see that getting good bounds on \(|\bias_\X(f)|\) for a random low-degree polynomial \(f\) is \emph{equivalent} to showing that a random low-degree polynomial is a good extractor for \(\X\). Thus, in order to figure out interesting distributions \(\X\) for which to pursue upper bounds on \(|\bias_\X(f)|\), it is only natural to borrow some motivation from extractor theory. We do so, below.

\subsubsection*{Key questions}
In order to show that a random low-degree polynomial extracts from a distribution \(\X\sim\zo^n\) (equivalently, in order to upper bound \(|\bias_\X(f)|\)), it is easy to see that an absolute minimum requirement is that \(\X\) contains some ``randomness.'' To formalize this notion, it is standard to use \emph{min-entropy}, defined as follows.

\[                          
    H_\infty(\X):=\min_{x\in\supp(\X)}\log\left(\frac{1}{\Pr[\X=x]}\right).
\]
If \(\X\) has min-entropy \(H_\infty(\X)\geq k\), we often refer to it as an \((n,k)\)-source.

Unfortunately, a well-known impossibility result says that even if each source \(\X\in\mathcal{X}\) has nearly \emph{full} min-entropy, it is still impossible to extract \cite{CG88}.\footnote{Crucially, this is because the extractor \(\Ext\) must be a single function that works \emph{for all} \(\X\in\mathcal{X}\).} Thus, in order to make extraction possible, we not only need a lower bound on the min-entropy of each \(\X\in\mathcal{X}\), but we also need to assume that each \(\X\in\mathcal{X}\) has some \emph{structure}. Towards this end, the oldest trick in the book is to assume that the family \(\mathcal{X}\) is not too large. In this case, since a \emph{uniformly random function} extracts from one source \(\X\) (with a min-entropy guarantee) with extremely high probability \cite[Proposition 6.12]{Vad12}, a simple union bound allows one to conclude that there exists a single function that extracts from \emph{all} sources \(\X\in\mathcal{X}\).

Given the above discussion, it is natural to ask whether an analogous fundamental result can be established for uniformly random \emph{low-degree polynomials}, which immediately raises the question,
\begin{center}
    \emph{How biased is a random degree \(d\) polynomial on a single \((n,k)\)-source \(\X\)?}
\end{center}
If we can show that \(|\bias_\X(f)|\) is low with extremely high probability over \(f\), then we can conclude that random low-degree polynomials extract from any small family \(\mathcal{X}\) of sources. Importantly, many well-studied families of sources are, in fact, very small. In particular, this is true of \emph{all families} for which random low-degree polynomial extractors have been studied: uniform sources \cite{BHL12}, affine sources \cite{CT15}, and local sources \cite{ACGLR22}.\footnote{The family of uniform sources is the trivial family \(\mathcal{X}=\{\U_n\}\), while the family of affine sources (of min-entropy \(k\)) consists of all \(\X\sim\F_2^n\) that are uniform over a \(k\)-dimensional affine subspace of \(\F_2^n\). Local sources, on the other hand, consist of sources of the form \(\X=f(\U_\ell)\), where \(f\) is some function where each output bit depends on a bounded number of input bits.} Thus, showing that a random low-degree polynomial is unbiased on any single source \(\X\) of min-entropy \(k\) could lead to a result that subsumes (and greatly generalizes) all previous work.

Unfortunately, several important families of sources \(\mathcal{X}\) are quite large. For these, the above approach cannot be used to show that random low-degree polynomials extract. Here, the most general (well-studied) family is the family \(\mathcal{X}\) of \emph{sumset sources} \cite{CL16}, a model inspired by fundamental structures in additive combinatorics. Formally, an \((n,k)\)-sumset source is defined to have the form \(\X=\A+\B\), where \(\A,\B\sim\zo^n\) are independent sources of min-entropy at least \(k\), and \(+\) denotes bitwise XOR. Sumset sources generalize a huge number of other well-studied large families \cite{CL16,chattopadhyay2022improved,CL22}, including: independent sources \cite{CG88}, interleaved sources \cite{RY11}, and small-space sources \cite{KRVZ11} (and affine sources \cite{ben2001extractors}, though this family is small). Thus, to complement our first question, we also ask:
\begin{center}
    \emph{How good is a random degree \(d\) polynomial as an extractor\\for the family of \((n,k)\)-sumset sources?}
\end{center}

In the remainder of this paper, our goal is to answer both of the questions presented above. In doing so, we hope to provide new insight into the power of a fundamental computational model (low-degree polynomials) for a fundamental computational task (randomness extraction). As it turns out, however, there are a few other reasons why answering these questions might be useful. Before we formally present our main results, we highlight some of these, below.

\paragraph{Pseudorandomness}

Low-complexity extractors have important applications in the real world and theory. In the real world, low-complexity extractors are more likely to be implemented and exhibit a reasonable running time. In theory, low-complexity extractors serve as fundamental building blocks in the construction of key cryptographic \cite{lu02,Vad04} and pseudorandom primitives \cite{Li11,li2023explicit}.\footnote{In fact, several well-known explicit extractors are \emph{themselves} low-degree polynomials \cite{CG88,BNS92,ben2001extractors,Bou05,Dvi09,LZ19}! For example, the inner-product extractor~\cite{CG88,ben2001extractors} is a degree $2$ polynomial of its input, while the generalized inner-product extractor~\cite{BNS92,Dvi09,LZ19} is its natural extension to degree $d$. A version of Bourgain's two-source extractor~\cite{Bou05} can also be seen as a low-degree polynomial over $\F_3$. More precisely, the version of Bourgain's extractor $\mathsf{Bou}(x,y)=\langle (x,x^2),(y,y^2)\rangle$, where $\langle\cdot,\cdot\rangle$ denotes the inner product over $\F_3$, with $x,y\in\F_3^n$ for $n$ prime and $x^2$ and $y^2$ computed in $\F_{3^n}$ (see~\cite[Section 3.5.2]{Dvi12}), is a degree $\leq 4$ $\F_3$-polynomial, since the representations in $\F_3^n$ of $x^2$ and $y^2$ can be computed as $n$-variate $\F_3$-polynomials of $x$ and $y$ of degree $\leq 2$~\cite[Lemma 2.3.1]{Kop10}.} Their study has also led to important structural results for well-studied families of distributions \cite{CT15,ACGLR22}. Because of this, low-complexity extractors have received a lot of attention in the literature \cite{dm02,lu02,Vad04,vio05,dt09,Li11,BG13,GVW15,CT15,CL18,ACGLR22,HIV22,CW24,li2023explicit}, with the works of Cohen and Tal \cite{CT15} and Alrabiah, Chattopadhyay, Goodman, Li, and Ribeiro \cite{ACGLR22} specifically focusing on the power of \emph{random low-degree polynomials} as extractors (for affine sources and local sources, respectively).

\paragraph{Coding theory}

Low-degree polynomials are fundamental objects in both algebra and coding theory, and studying whether they are good extractors ultimately requires proving new structural results about them (leading to new insights in these two areas). As an example, the work of Ben-Eliezer, Hod, and Lovett \cite{BHL12} (on low-degree extractors for uniform sources) immediately gave new bounds on the weight distribution of Reed-Muller codes. On the other hand, the results of Cohen and Tal \cite{CT15} (on low-degree extractors for affine sources) showed that every low-degree polynomial must have a big subspace in its solution set. And the paper of Alrabiah, Chattopadhyay, Goodman, Li, and Ribeiro \cite{ACGLR22} (on low-degree extractors for local sources) proved a new type of ``Chevalley-Warning theorem,'' which established that every (small) system of low-degree polynomial equations must have a solution with low Hamming weight.

\paragraph{Complexity theory}

Finally, low-complexity extractors can help us establish fine-grained complexity separations (as advocated for in, e.g., \cite{hatami2023depth}). In more detail, extractors are known to exhibit strong lower bounds against a variety of well-studied complexity classes, including low-depth circuits \cite{hrubes2015circuits,GKW21}, general circuits \cite{demenkov2011elementary,golovnev2016weighted,find2016better,GKW21,li20221}, various flavors of branching programs \cite{hrubes2015circuits,ourSpaceComplexity,gryaznov2022linear,CL23}, and more \cite{cohen2016complexity,GKW21,GG25,CG25}. Showing that there exist extractors in a low-level complexity class \(\mathcal{C}\) would allow one to separate \(\mathcal{C}\) from the classes above.

\subsection{Our results}

In this paper, we show that random low-degree polynomials extract from any small family of sources, and from the (large) family of sumset sources. This answers both questions presented in the introduction, and we present these two results (respectively) in \cref{subsubsec:small-families,subsubsec:sumset-sources}.

\subsubsection{Low-degree polynomials extract from small families}\label{subsubsec:small-families}

In order to prove that random low-degree polynomials can extract from any small family, we first show that a random low-degree polynomial extracts from a single source. In particular, we prove the following, which can be viewed as both (1) a ``low-degree'' version of the classical fact that a random \emph{function} extracts from a single source \cite[Proposition 6.12]{Vad12}, and (2) a generalization of the result that a random low-degree polynomial has low bias  \cite[Lemma 1.2]{BHL12} (\cref{intro:thm:BHL12}).

\begin{restatable}[Low-degree polynomials extract from a single source]{theorem}{weaksource}\label{our-results:main:first-theorem}
    For every \(\delta\in(0,1)\) there is a constant \(c>0\) such that the following holds. Let \(\X\sim\F_2^n\) be a source with min-entropy at least \(k\), and let \(d\in\N\) be an integer satisfying \(1\leq d\leq(1-\delta)k\). Then for a random degree \(d\) polynomial \(f:\F_2^n\to\F_2\),
    \[
        \Pr_f\left[|\bias_\X(f)|>2^{-ck/d}\right]\leq2^{-c\binom{k}{\leq d}}.
    \]
\end{restatable}

We highlight some key aspects of this result. First, it has a simple proof, which follows by combining \cite[Lemma 1.2]{BHL12} (\cref{intro:thm:BHL12}) with the leftover hash lemma \cite{HILL99}. Second, it is easy to verify that it is tight.\footnote{This follows by considering the \((n,k)\)-source \(\X\) which is uniform on the first \(k\) bits and constantly \(0\) on the remaining bits, combined with the tightness of \cref{intro:thm:BHL12}.} Third, we emphasize that the above result works for \emph{any} distribution of min-entropy at least \(k\), not just those that are ``flat'' (uniform over a subset \(S\subseteq\F_2^n\) of size \(2^k\)). This is crucial in some applications.\footnote{Even though arbitrary \((n,k)\)-sources are known to be convex combinations of flat \((n,k)\)-sources \cite[Lemma 6.10]{Vad12}, this convex combination may end up bringing the source \(\X\) out of the ``small family'' \(\mathcal{X}\).}

We also note that by a standard application of the XOR lemma \cite[Lemma 3.8]{dodis2014privacy}, it is straightforward to extend \cref{our-results:main:first-theorem} to show that a \emph{sequence} of independent, uniformly random degree \(d\) polynomials \(f_1,f_2,\dots,f_m:\F_2^n\to\F_2\) can be concatenated to create a multi-bit extractor for \(\X\).\footnote{In order to apply the XOR lemma, the only observation needed is that the XOR of (any number of) independent, uniformly random degree \(d\) polynomials applied to \(\X\) is, itself, a uniformly random degree \(d\) polynomial applied to \(\X\).} In fact, this can further be extended to show that the sequence \(f_1,f_2,\dots,f_m\) not only extracts \(m\) uniform bits from \(\X\), but has low correlation with any (short) fixed function \(g\) applied to \(\X\).\footnote{This can be done by first conditioning on the output of \(g(\X)\), which will only cause \(\X\) to lose a little bit of min-entropy.} Finally, if we set \(k=n\), \(m=1\), and \(g\) to have output length \(1\), this result can be interpreted as nontrivial bounds on the list-size of Reed-Muller codes at the extreme (relative) radius of \(1/2-2^{-\Omega(n/d)}\). This appears to be the first result of this form (c.f.\ \cite{kaufman2012weight,ASW15,abbe2020reed}), and naturally extends to punctured Reed-Muller codes (by picking \(k<n\)).

Returning to our original problem, it is straightforward to combine \cref{our-results:main:first-theorem} to show our unifying result: that random low-degree polynomials extract from any small family of sources, with exponentially small error. We record this corollary below, and instantiate the general result with three interesting small families of sources: local sources, polynomial sources, and variety sources. The family of local sources are easily shown to be small, while the families of polynomial sources and variety sources were recently shown to be small via ``input reduction lemmas'' \cite{CGG24,CT15}.

\begin{corollary}[Low-degree polynomials extract from small families]\label{coro:smallfamilies}
For every \(\delta>0\) there exists a constant \(c>0\) such that for all $n\geq k\geq d\in\N$ with $d\leq (1-\delta)k$, the following holds. 
For any family \(\mathcal{X}\) of \((n,k)\)-sources with size \(|\mathcal{X}|<2^{c\binom{k}{\leq d}}\), a random degree \(d\) polynomial \(f:\zo^n\to\zo\) is a \(2^{-ck/d}\)-extractor for \(\mathcal{X}\), except with probability at most \(2^{-c\binom{k}{\leq d}}\).

In particular, we obtain the following for a sufficiently large constant $C>0$ depending only on $\delta$.

\begin{itemize}
    \item \textbf{Local sources}: There exists a degree \(\leq d\) polynomial \(f:\zo^n\to\zo\) that is a \(2^{-ck/d}\)-extractor for \(r\)-local sources with min-entropy
    \[
    k\geq Cd(2^r n+r n\log n)^{1/d}.
    \]

    \item \textbf{Polynomial sources}: There exists a degree \(\leq d\) polynomial \(f:\zo^n\to\zo\) that is a \(2^{-ck/d}\)-extractor for degree \(r\) polynomial sources with min-entropy
    \[
    k\geq C\left(\frac{C^{r}d^d}{r^{r}}\cdot n\right)^{1/(d-r)}.
    \]    
    \item \textbf{Variety sources}: There exists a degree \(\leq d\) polynomial \(f:\zo^n\to\zo\) that is a \(2^{-ck/d}\)-extractor for degree \(r\) variety sources with min-entropy
    \[
    k\geq Cdn^{(r+1)/d}.
    \]
\end{itemize}
\end{corollary}

We make a few brief remarks about this result. First, we note that our result on local sources significantly improves the parameters of the previous best result \cite[Theorem 1.1]{ACGLR22}, which required min-entropy \(k\geq C2^rr^2d(2^rn\log n)^{1/d}\) and had error \(\eps=2^{-ck/(d^32^rr^2)}\). Second, we highlight that our result on polynomial sources may be surprising, as it shows that polynomials can be used to extract from polynomial sources. Perhaps this can be used to make progress on (the challenging goal of) constructing explicit extractors for polynomial sources \cite{DGW09,CGG24}, as it suggests that it is possible to improve the quality of a polynomial source \emph{while keeping it as a polynomial source}. Third, we mention one interesting application of our result on variety sources. By exploiting known hardness results for variety extractors \cite{li20221,golovnev2016weighted,GKW21}, it follows that (general) circuits of size \(3.1n-o(n)\) cannot compute (all) degree \(3\) polynomials, circuits of size \(3.11n-o(n)\) cannot compute degree \(4\) polynomials, and circuits of size \(3.9n-o(n)\) cannot compute degree \(18\) polynomials. Finally, we note that in \cref{sec:seeded}, we show that \cref{coro:smallfamilies} (in fact, \cref{intro:thm:BHL12}) can be used to prove low-degree polynomials are good \emph{linear seeded extractors}.\footnote{Recall that a linear seeded extractor only needs to be linear on the source (fixing the seed), but may be an arbitrarily high-degree polynomial in general.}

\paragraph{Concurrent work} In a concurrent and independent work, Golovnev, Guo, Hatami, Nagargoje, and Yan \cite{GGHNY24} prove similar results to those presented above (in \cref{subsubsec:small-families}). In particular, they show that random low-degree polynomials are good extractors for any small family of sources, and instantiate this to obtain results similar to those presented in \cref{coro:smallfamilies}. Moreover, our proofs both rely on a similar key ingredient on the dimension of punctured Reed-Muller codes \cite[Theorem 1.5]{KS05}. The differences are as follows: our result achieves better error (exponentially small vs.\ polynomially small), and we also establish results for sumset sources (discussed in \cref{subsubsec:sumset-sources} below). On the other hand, the work \cite{GGHNY24} proves a significant generalization of \cite[Theorem 1.5]{KS05}, in order to get interesting results in algebraic geometry.

\subsubsection{Low-degree polynomials extract from sumset sources}\label{subsubsec:sumset-sources}

In the second main part of our paper, we show that random low-degree polynomials are also good extractors for \emph{sumset sources}, which are the most general well-studied \emph{large} family of sources. 

\begin{restatable}[Low-degree polynomials extract from sumset sources]{theorem}{lowdegsumsetext}\label{thm:lowdegsumsetext}
There exists a constant \(C>0\) such that for any \(n\geq k\geq d\in\N\) and \(\eps>0\) such that \(k\geq Cd(n/\eps^2)^{1/\lfloor d/2\rfloor}\), a random degree \(d\) polynomial \(f:\F_2^n\to\F_2\) is an \(\eps\)-extractor for \((n,k)\)-sumset sources, with probability at least \(1-2^{-\eps^2\binom{k/C}{2\lfloor d/2\rfloor}}\geq1-2^{-n^2/\eps^2}\).
\end{restatable}

We highlight a few key specializations of the above theorem -- focusing on even \(d\), for simplicity. First, we note that in the \emph{disperser} regime,\footnote{A \emph{disperser} is an extractor \(f:\F_2^n\to\F_2\) with nontrivial error \(\eps<1/2\).} \cref{thm:lowdegsumsetext} shows that there exist degree \(\leq d\) polynomials that disperse from sumset sources with min-entropy \(k=O(dn^{2/d})\). This is nearly tight, since Cohen and Tal \cite{CT15} show that degree \(\leq d\) polynomials cannot extract from \emph{affine sources} (and thus sumset sources) with min-entropy below \(k=\Omega(dn^{1/(d-1)})\). (In \cref{sec:impossibility}, we show the same impossibility result holds for \emph{independent sources}, which are the other special case of sumset sources.) Second, we note that in the \emph{polynomial error regime} \(\eps=n^{-\gamma}\), \cref{thm:lowdegsumsetext} shows that there exist degree \(\leq d\) polynomials that are \(\eps\)-extractors for sumset sources with min-entropy \(k=O(dn^{2(1+2\gamma)/d})\). Third, in the \emph{arbitrary error regime} \(\eps>0\), \cref{thm:lowdegsumsetext} shows that a random degree \(d=O(\log(n/\eps))\) polynomial \(f\) is an \(\eps\)-extractor for sumset sources with min-entropy \(k=O(\log(n/\eps))\). This strengthens the existential result of Mrazovi\'c \cite{Mra16}, who obtained such a min-entropy requirement for a uniformly random function \(f\).

Finally, we highlight that our sumset extractors have interesting consequences for \emph{linear read-once branching programs}, and our proof requires two new tools, which may be of independent interest.

\paragraph{Linear ROBPs} In more detail, linear read-once branching programs (ROBPs) are a new type of computational model \cite{gryaznov2022linear}, which simultaneously generalize both standard ROBPs and parity decision trees. At each point in the branching program, instead of querying a single input \emph{variable}, the ROBP is allowed to query an arbitrary \emph{linear function} of the input (so long as it is linearly independent of all previous queries). We observe that by leveraging standard results on finite fields \cite[Lemma 6.21]{finiteFieldsBook} (see also \cite[Lemma 17]{bogdanov2010pseudorandom}), linear ROBPs of constant width \(w=O(1)\) can compute any polynomial of degree \(2\).\footnote{In more detail, \cite[Lemma 17]{bogdanov2010pseudorandom} asserts that for any quadratic \(q:\F_2^n\to\F_2\), there is some \(B\in\F_2^{m\times n}\) with full row rank and some affine \(L:\F_2^n\to\F_2\) such that \(q(x)=\langle (Bx)_{\leq m/2}, (Bx)_{>m/2} \rangle + L(x)\). Since each row in \(B\) is linearly independent, the inner product can be computed using \(\leq 2\) bits of storage. Then, using \(\leq1\) additional bit of storage (in case \(L\) is linearly dependent on the rows in \(B\)), one can simultaneously compute \(L(x)\). Thus, \(q(x)\) can be computed by a constant-width linear ROBP.} On the other hand, \cref{thm:lowdegsumsetext} (combined with \cite[Theorem 1]{CL23}) implies that linear ROBPs require \emph{exponential} width \(w=2^{n-o(n)}\) to compute polynomials of degree \(4\). This is a huge, perhaps surprising, jump in complexity.\footnote{Indeed, one might expect the width \(w\) to somehow grow proportionately with the degree \(d\) of the polynomial that must computed. However, this shows that the width jumps from constant to \emph{exponential}, simply by moving from degree \(2\) to degree \(4\).}

\paragraph{Sumset-punctured Reed-Muller codes} Finally, we highlight that the proof of \cref{thm:lowdegsumsetext} requires two new key ingredients, which may be of independent interest. The first key ingredient is a new result about the structure of Reed-Muller codes that are punctured on sumsets. As a bonus application, we use this to build new ``evasive sets,'' which we then use to improve the extractor impossibility results of Chattopadhyay, Goodman, and Gurumukhani \cite{CGG24}, and to get a low-error version of \cref{thm:lowdegsumsetext} for the special case of degree \(4\) polynomials and independent sources. Interestingly, the latter result relies on the recent breakthrough resolution of Marton's \emph{PFR conjecture} from additive combinatorics \cite{GGMT23}.

\paragraph{A novel reduction between extractors} The second key ingredient in the proof of \cref{thm:lowdegsumsetext} is a novel type of reduction between extractors. While most reductions between extractors rely on showing that a source can be equipped with structure by breaking it down into a convex combination of well-behaved distributions via a \emph{deterministic process}, we show that doing so via a careful \emph{(correlated) randomized process} can sometimes make this task much easier. 
In \cref{sec:randsumsetext}, we illustrate a simpler variant of this idea in order to give an alternative proof that a uniformly random function is an extractor for sumset sources with min-entropy \(k=O(\log(n/\eps))\) -- a result first established by Mrazovi\'{c} \cite{Mra16}.

\subsubsection*{Organization}

The remainder of the paper is organized as follows. In \cref{sec:overview}, we provide an overview of our main techniques. Then, in \cref{sec:prelims}, we record some basic preliminaries on notation, probability, and extractors. The first main technical part of this paper is \cref{sec:smallfamilies}, where we show that random low-degree polynomials extract from small families, and prove all results listed in \cref{subsubsec:small-families}. The second main technical part of the paper consists of \cref{sec:largefamilies}, where we prove our main result from \cref{subsubsec:sumset-sources}. In \cref{sec:evasive}, we show that one of the key ingredients from the prior section can be used to establish the existence of ``evasive sets,'' which are then used to obtain a few additional results. (In \cref{sec:randsumsetext}, we show that the other key ingredient can be used to obtain an alternative proof of the result by Mrazovi\'{c} \cite{Mra16}.) Finally, we wrap up with some open problems in \cref{sec:conclusions}.

\dobib

\section{Overview of our techniques}\label{sec:overview}

\subsection{Low-degree polynomials extract from small families}

Recall that our goal in \cref{our-results:main:first-theorem} is to obtain concentration bounds for $|\bias_\X(f)|=|\E_{x\sim \X}[(-1)^{f(x)}]|$ with $\X$ an arbitrary $(n,k)$-source.
Our simple argument combines the original result of~\cite{BHL12} for $\X$ uniformly distributed over $\F_2^n$ (\cref{intro:thm:BHL12}) and an application of the leftover hash lemma.

We first introduce some useful concepts.
For a vector $x\in\F_2^n$, we denote by $\eval_d(x)$ the tuple of evaluations of all monomials of degree at most $d$ on $x$, i.e.,
\begin{equation*}
    \eval_d(x)=\left(\prod_{i\in I}x_i\right)_{I\subseteq [n], |I|\leq d}\in\F_2^{\binom{n}{\leq d}}.
\end{equation*}
Given a set $S\subseteq\F_2^n$, we write $\eval_d(S)=\{\eval_d(x):x\in S\}$.

In order to obtain the desired concentration, it suffices to appropriately upper bound the high-order moments $\E[\bias_\X(f)^t]$ for a large $t$, which is also the approach followed in~\cite{BHL12}.
And, also analogously to~\cite{BHL12}, it is not hard to show that
\begin{align}
    \E_f[\bias_\X(f)^t] &= \Pr_{x^{(1)},\dots,x^{(t)}\sim \X}[\eval_d(x^{(1)})+\cdots+\eval_d(x^{(t)})=0]\nonumber\\
    &=\Pr_{x^{(1)},\dots,x^{(t)}\sim \X}[\forall p:\F_2^n\to\F_2,\deg p\leq d:p(x^{(1)})+\cdots+p(x^{(t)})=0].\label{eq:biastPoly}
\end{align}

Intuitively, we would like to reduce the task of bounding $\E_f[\bias_\X(f)^t]$ to the task of bounding $\E_f[\bias(f)^t]$, which can be handled via the concentration bounds from~\cite{BHL12}.
We establish such a reduction via the leftover hash lemma, which guarantees the existence of a \emph{linear} map $L:\F_2^n\to\F_2^m$ with $m\approx k$ such that $L(\X)$ is close (in statistical distance) to the uniform distribution on $\F_2^m$.
We claim that
\begin{equation}\label{eq:redlinmap}
    \E_f[\bias_\X(f)^t]\leq \E_g[\bias_{L(\X)}(g)^t]
\end{equation}
where $g:\F_2^m\to\F_2$ is a random degree $d$ polynomial. This holds since, by \cref{eq:biastPoly},
\begin{align*}
    \E_g[\bias_{L(\X)}(g)^t] &= \Pr_{y^{(1)},\dots,y^{(t)}\sim L(\X)}[\forall q:\F_2^m\to\F_2,\deg q\leq d:q(y^{(1)})+\cdots+q(y^{(t)})=0]\\
    &=\Pr_{x^{(1)},\dots,x^{(t)}\sim \X}[\forall q:\F_2^m\to\F_2,\deg q\leq d:q(L(x^{(1)}))+\cdots+q(L(x^{(t)}))=0]\\
    &\geq \eqref{eq:biastPoly},
\end{align*}
where the inequality uses the fact that $q\circ L:\F_2^n\to\F_2$ has degree at most $d$
(with $q\circ L$ denoting composition), as $L$ is linear.

We are almost done.
Informally, since $L(\X)\approx \U_m$, it is easy to show that $\E_g[\bias_{L(\X)}(g)^t]\approx \E_g[\bias(g)^t]$.
Moreover, we can upper bound $\E_g[\bias(g)^t]$ appropriately via the known concentration bound from~\cite{BHL12}.
Combining this with \cref{eq:redlinmap} yields the desired upper bound on $\E_f[\bias_\X(f)^t]$, which we translate into a concentration bound on $|\bias_\X(f)|$ via Markov's inequality.

\subsection{Low-degree polynomials extract from sumset sources}\label{sec:TOsumsetExt}

Next, we discuss the approach behind \cref{thm:lowdegsumsetext}. For simplicity, we focus here on the case of even degree \(d\), and note that it is trivial to extend to odd \(d\) (at a slight loss in parameters).

\subsubsection{Low-degree polynomials disperse from sumset sources}\label{sec:TOsumsetDisp}

As a warm-up, we first consider the simpler task of \emph{dispersing}. In this case, we wish to show that a random degree $d$ polynomial $f:\F_2^n\to\F_2$ will satisfy $f(\supp(\W))=\zo$ simultaneously for all $(n,k)$-sumset sources $\W$ with $k=O(dn^{2/d})$.
Then, we discuss the necessary modifications to obtain sumset extraction with arbitrary error \(\eps>0\).

First, as usual, we only need to focus on sumset sources $\W=\X+\Y$ where $\X$ and $\Y$ are independent \emph{flat} $(n,k)$-sources.
Denote the supports of $\X$ and $\Y$ by $X$ and $Y$, respectively, which have size $2^k$. Then, the probability that $f$ is identically $0$ on $X+Y$ is
\begin{equation}\label{eq:UBrankProb0}
    \Pr_f[f(X+Y)\equiv 0]\leq 2^{-\rank(\eval_d(X+Y))}.
\end{equation}
This is because we may write $f(x)=\langle v, \eval_d(x)\rangle$ for a uniformly random vector $v\in\F_2^{\binom{n}{\leq d}}$, and so (i) $f(x)$ is uniformly distributed over $\F_2$ for any fixed nonzero $x$, and (ii) $f(x)$ and $f(y)$ are independent (and uniformly distributed) if $\eval_d(x)$ and $\eval_d(y)$ are linearly independent.

Given \cref{eq:UBrankProb0}, it is clear that we must understand $\rank(\eval_d(X+Y))$.
If this quantity is suitably large, then the probability that $f$ is constant on any such sumset $X+Y$ is small, and we could hope to survive a union bound over all choices of $X$ and $Y$.
However, this strategy cannot directly work, because $\rank(\eval_d(X+Y))$ will be at most $\binom{n}{\leq d}$ while there are $\binom{2^n}{2^k}^2\geq 2^{2(n-k)2^k}\gg 2^{\binom{n}{\leq d}}$ choices for $X$ and $Y$.

A possible way to overcome the barrier to the application of the union bound above is to show that there exist appropriately small subsets $X'\subseteq X$ and $Y'\subseteq Y$ such that $\rank(\eval_d(X'+Y'))$ is still large.
If this were the case, we could then just apply the union bound over all possible choices of the now much smaller sets $X'$ and $Y'$.
We can make this approach work by proving the following:

\begin{claim}[Informal]\label{claim:key1}
    Let $A,B\subseteq\F_2^n$ be sets of size $2^k$.
    Then, there exist subsets $A'\subseteq A$ and $B'\subseteq B$ each of size roughly $\sqrt{\binom{k}{\leq d}}$ such that $\rank(\eval_d(A'+B'))$ is roughly $\binom{k}{\leq d}$.
\end{claim}

We sketch how \cref{claim:key1} can be applied to obtain the desired result.
Setting $A=X$ and $B=Y$, we obtain $X'\subseteq X$ and $Y'\subseteq Y$ of size about $\sqrt{\binom{k}{\leq d}}$ such that $\eval_d(X'+Y')$ has rank about $\binom{k}{\leq d}$.
By \cref{eq:UBrankProb0}, this means that the probability that $f$ is identically $0$ on $X'+Y'$, and hence on $X+Y$, is at most about $2^{-\binom{k}{\leq d}}$.
But now, crucially, we only have to carry out a union bound over the at most about $\binom{2^n}{k^{d/2}}^2$ choices for $X'$ and $Y'$, which is approximately $2^{n k^{d/2}}\ll 2^{\binom{k}{\leq d}}$ when $k\geq Cd n^{2/d}$ for some large enough constant \(C>0\). Note that this strategy would not have worked if $X'$ and $Y'$ were instead of size close to $\binom{k}{\leq d}$.

Before discussing the simple proof of \cref{claim:key1}, it is instructive to consider existing results of a similar flavor.
Keevash and Sudakov~\cite{KS05} (and later Ben-Eliezer, Hod, and Lovett~\cite[Lemma 1.4]{BHL12}) proved that for any set $S\subseteq\F_2^n$ of size $2^k$ there exists a subset $S'\subseteq S$ of size at least $\binom{k}{\leq d}$ such that $\rank(\eval_d(S'))=|S'|$.
This lemma is one of the most important steps in the proof of the main result of~\cite{BHL12}, and its proof hinges on the construction of an intricate ``rank-preserving surjection'' from $S$ to $\F_2^k$.
More precisely, this is a map $g:\F_2^n\to\F_2^k$ that is surjective on $S$ and satisfies
\begin{equation*}
    \rank(\eval_d(S)) \geq \rank(\eval_d(g(S))) = \rank(\eval_d(\F_2^k))=\binom{k}{\leq d}.
\end{equation*}
Applying this same rank-preserving surjection in the setting of \cref{claim:key1} is not guaranteed to work, because now the subset that witnesses the $\binom{k}{\leq d}$ rank lower bound must be a sumset $A+B$ with $|A|,|B|\approx \sqrt{\binom{k}{\leq d}}$.
We overcome this by observing that if we are fine with worse constants, then we can replace the complicated rank-preserving surjection from~\cite{KS05,BHL12} with a \emph{linear} map, guaranteed by the leftover hash lemma.
In fact, by the leftover hash lemma, there exists a linear map $L:\F_2^n\to\F_2^{k'}$, for $k'=\Omega(k)$, which is surjective on both $X$ and $Y$.
We choose $X'\subseteq X$ and $Y'\subseteq Y$ such that $L(X')=L(Y')=\cB^{k'}_{d/2}(0)$, the radius-$d/2$ Hamming ball in $\F_2^{k'}$ centered at $0$.
Then, the linearity of $L$ plus basic properties of evaluation vectors allows us to conclude that
\begin{align*}
    \rank(\eval_d(X'+Y'))&\geq \rank(\eval_d(L(X'+Y')))\\
    &=\rank(\eval_d(\cB^{k'}_{d/2}(0)+\cB^{k'}_{d/2}(0)))=\rank(\eval_d(\cB^{k'}_d(0))),
\end{align*}
and it is well known that $\rank(\eval_d(\cB^{k'}_d(0)))=\binom{k'}{\leq d}$.

\subsubsection{From dispersers to extractors via random convex combinations}\label{sec:TOfromDispToExt}

The argument discussed above shows that a random degree $d$ polynomial is a $k$-sumset disperser for min-entropy $k=O(dn^{2/d})$ with high probability.
It remains to see how we can extend this to get sumset extraction for similar min-entropy $k$ with arbitrary error $\eps>0$.

Our first observation is that if $\W=\X+\Y$ is a sumset source with flat $\X$ and $\Y$ whose supports $X$ and $Y$ satisfy $\rank(\eval_d(X+Y))=|X|\cdot |Y|$, 
then $\Pr[f(\W)=0]\approx 1/2$ holds with high probability over the choice of a random degree $d$ polynomial $f$.
In other words, $f(\W)$ is close (in statistical distance) to uniformly distributed over $\zo$ with high probability over the choice of $f$.\footnote{The exact shape of the ``high probability'' comes from a standard Chernoff bound.} This happens because, under the conditions above, $\W$ is a flat source and all the vectors in $\eval_d(X+Y)$ are linearly independent, which means that the bits $(f(w)=\langle v,\eval_d(w)\rangle)_{w\in X+Y}$ are independent and uniformly distributed when $v$ is a uniformly random vector over $\F_2^{\binom{n}{\leq d}}$.

Now, set $k=Cd (n/\eps^2)^{2/d}$ for a large enough constant \(C>0\), and call a sumset source $\W=\X+\Y$ \emph{special} if $\X$ and $\Y$ are flat, $\rank(\eval_d(X+Y))=|X|\cdot |Y|$, and $|X|,|Y|\approx \sqrt{\binom{k}{\leq d}}$. Combining the previous paragraph with a union bound over the choices of $X$ and $Y$ (analogous to the one in \cref{sec:TOsumsetDisp}) shows that a random degree $d$ polynomial will be, with high probability, an $\eps$-extractor for the class of special sumset sources.

Of course, most sumset sources are far from special.
We overcome this by showing that every $(n,k)$-sumset source $\W=\X+\Y$ with flat $\X$ and $\Y$ is $2^{-\Omega(k)}$-close to a convex combination of special sumset sources.
Combining this with the observations above lets us conclude that a random degree $d$ polynomial will be a $(k,\eps'=2^{-\Omega(k)}+\eps \approx \eps)$-sumset extractor with high probability.

It remains to argue why every $(n,k)$-sumset source $\W=\X+\Y$ with flat $\X$ and $\Y$ is $2^{-\Omega(k)}$-close to some convex combination of special sumset sources.
To better highlight the main underlying ideas, we consider here only the particular case where $k=n$ and $\X$ and $\Y$ are uniformly distributed over $\F_2^k$, and present a less optimized version of our final argument.
It is not hard to reduce the general case to a scenario very similar to this particular case through an application of the leftover hash lemma.

We consider an alternative way of (approximately) sampling from $\W=\X+\Y$ -- the convex combination will be implicit in this sampling procedure.
The idea is to first sample uniformly random subsets $X'\subseteq X=\F_2^k$ and $Y'\subseteq Y=\F_2^k$ each of size $\binom{k/3}{d/2}$ (which is \emph{very roughly} $\sqrt{\binom{k}{\leq d}}$), and then sample $\X'$ and $\Y'$ uniformly at random from $X'$ and $Y'$, respectively.
If $X'$ and $Y'$ are sampled independently, then it is not hard to show that $\W'=\X'+\Y'$ is distributed exactly like $\W$.
However, we would like to claim that the resulting sumset source $\X'+\Y'$ will be special with high probability over the choice of subsets $X'$ and $Y'$.
To this end, we do not sample the subsets $X'$ and $Y'$ independently from each other, but rather couple the randomness used in their sampling carefully.
This coupling will ensure that $\W'=\X'+\Y'$ is always a special sumset source for any fixing of $X'$ and $Y'$, while we still have $\W'\approx_{2^{-\Omega(k)}}\W$.

To sample the (\emph{correlated}) random subsets $X'$ and $Y'$, we proceed as follows.
Let $B_1=\{u_1,\dots,u_t\}$ be the set of weight-$d/2$ vectors supported on $\{1,\dots,k/3\}$, and let $B_2=\{v_1,\dots,v_t\}$ be the set of weight-$d/2$ vectors supported on $\{2k/3+1,\dots,k\}$.
Since any two vectors $u_i$ and $v_j$ have disjoint supports and are non-zero, each sum $u_i+v_j$ is a distinct non-zero vector in the radius-$d$ Hamming ball, and so
\begin{equation*}
    \rank \eval_d(B_1+B_2) = |B_1|\cdot |B_2| = \binom{k/3}{d/2}^2.
\end{equation*}
We couple the sampling of $X'$ and $Y'$ by choosing a uniformly random \emph{invertible} matrix $\mathbf{L}\in\F_2^{k\times k}$ and setting $X'=\mathbf{L}B_1 = \{\mathbf{L}u_1,\dots,\mathbf{L}u_t\}$ and $Y'=\mathbf{L}B_2 = \{\mathbf{L}v_1,\dots,\mathbf{L}v_t\}$.
Now, because $\mathbf{L}$ is invertible, we know that
\begin{equation*}
    \rank \eval_d(X'+Y') = \rank \eval_d(B_1+B_2) =|B_1|\cdot |B_2|= |X'|\cdot |Y'|.
\end{equation*}
Therefore, if $\X'$ and $\Y'$ are sampled independently and uniformly at random from $X'$ and $Y'$, respectively, then $\W'=\X'+\Y'$ is a special sumset source, as desired.

However, because the choices of $X'$ and $Y'$ are now correlated, we still need to argue that this overall sampling process produces something statistically close to $\W=\X+\Y$, with $\X$ and $\Y$ independent and uniformly distributed over $\F_2^k$.
If $\mathbf{L}\in\F_2^{k\times k}$ was a uniformly random matrix, then this would be immediate.
Indeed, let $I$ and $J$ be the random indices associated to the choices of $\X'$ and $\Y'$ from $X'$ and $Y'$.
Then,
$\X'=\mathbf{L} u_I$ and $\Y'=\mathbf{L} v_J$ would be independent and uniformly distributed over $\F_2^k$, since $u_I$ and $v_J$ are linearly independent for all choices of $I$ and $J$.
To argue that this is still approximately true when $\mathbf{L}$ is required to be invertible, we use the fact that the supports of $u_I$ and $v_J$ lie in a subset of $2k/3$ coordinates.
Therefore, it suffices to focus on $2k/3$ columns of $\mathbf{L}$.
Since a collection of $2k/3$ uniformly random vectors over $\F_2^k$ will be linearly independent except with probability $2^{-\Omega(k)}$, we conclude that any collection of $2k/3$ columns of $\mathbf{L}$ will be $2^{-\Omega(k)}$-close in statistical distance to a collection of $2k/3$ uniformly random vectors.
This gives that $(\X',\Y')\approx_{2^{-\Omega(k)}} (\X,\Y)$, where $\X$ and $\Y$ are independent, and so $\W'=\X'+\Y'$ is $2^{-\Omega(k)}$-close to the true sumset $\W=\X+\Y$.

\dobib

\section{Preliminaries}\label{sec:prelims}

\paragraph{Notation}

We denote random variables by boldfaced uppercase letters such as $\X$ and $\Y$ and denote sets by uppercase letters such as $A$ and $B$ or, at times, by calligraphic uppercase letters.
In this work we focus on random variables supported on finite sets, and write $\supp(\X)$ for the support of the random variable $\X$.
We denote the uniform distribution over $\F_2^m$ by $\U_m$, and we write $\log$ for the base-$2$ logarithm. We use \(\mathsf{wt}(x)\) to denote the Hamming weight of a vector \(x\in\F_2^n\), and we let \(\mathcal{B}_r^n(v)\) denote the Hamming ball in \(\F_2^n\) that is centered at \(v\) and has radius \(r\). Finally, we define $\binom{n}{\leq r}=\sum_{i=0}^r\binom{n}{i}$.

\paragraph{Binomial coefficients} We will need the following fact about binomial coefficients.

\begin{fact}[\protect{\cite[Proposition 8]{BHL12}}]\label{lem:binom-large-t}
    For any $\beta,\delta\in(0,1)$, there exists a constant $\gamma>0$ such that the following holds for all $n\in\N$.
    If $1\leq d\leq \delta n$, $t\geq d$, and $t\geq (1-\gamma/d)n$, then it holds that $\binom{t}{\leq d}\geq \beta \binom{n}{\leq d}$.
\end{fact}

\subsection{Probability}

In this section we collect some basic notions from probability theory that will be useful throughout.

\begin{definition}[Statistical distance]
    The statistical distance between discrete random variables $\X$ and $\Y$ supported on $S$, denoted $\Delta(\X,\Y)$, is given by
    \begin{equation*}
        \Delta(\X,\Y)=\max_{T\subseteq S} \left|\Pr[\X\in T]-\Pr[\Y\in T]\right|=\frac{1}{2}\sum_{x\in S}\left|\Pr[\X=x]-\Pr[\Y=x]\right|.
    \end{equation*}
    We say that $\X$ and $\Y$ are \emph{$\eps$-close}, and write $\X\approx_\eps\Y$, if $\Delta(\X,\Y)\leq \eps$.
\end{definition}

We will heavily exploit the following standard result about statistical distance.

\begin{fact}[Data-processing inequality]
    For any random variables \(\X,\Y\sim V\) and function \(f:V\to W\),
    \[
    \Delta(\X,\Y)\geq\Delta(f(\X),f(\Y)).
    \]
\end{fact}

\begin{definition}[Min-entropy]
    The \emph{min-entropy} of a random variable $\X$ is defined as
    \begin{equation*}
    \minH(\X):=\min_{x\in\supp(\X)}\log\left(\frac{1}{\Pr[\X=x]}\right).
    \end{equation*}
\end{definition}

\begin{definition}[$(n,k)$-source]
    We say that $\X\sim\F_2^n$ is an \emph{$(n,k)$-source} if $\minH(\X)\geq k$.
\end{definition}

\begin{definition}[$(n,k)$-sumset source]
    We say that $\W\sim\F_2^n$ is an \emph{$(n,k)$-sumset source} if there exist independent $(n,k)$-sources $\X,\Y\sim\F_2^n$ such that $\W=\X+\Y$.
\end{definition}
\noindent We also use the following version of the Chernoff bound, which can be found in~\cite[Theorem 2.21]{Vad12}.
\begin{lemma}[Chernoff bound]\label{lem:chernoff}
    Let $\X_1,\dots,\X_t$ be independent random variables supported on $[0,1]$, and let $\X=\frac{1}{t}\sum_{i=1}^t \X_i$.
    Then,
    \begin{equation*}
        \Pr[|\X-\E[\X]|>\eps]\leq 2\cdot e^{-\eps^2 t/4}.
    \end{equation*}
\end{lemma}

Finally, we will need the following result from \cite{CG25}. We include its short proof, for completeness.

\begin{definition}
We say that \(g:Y\to X\) is a \emph{pseudoinverse} of \(f:X\to Y\) if \(f(g(f(x)))=f(x)\) for all \(x\in X\).
\end{definition}

\begin{lemma}[Dependency reversal \cite{CG25}]\label{lem:dependency-reversal}
For any random variable \(\X\sim X\) and deterministic function \(f:X\to Y\), there exists an independent random variable \(\A\sim A\) and deterministic function \(g:Y\times A\to X\) such that
\[
g(f(\X),\A)\equiv \X,
\]
and such that \(g(\cdot,a)\) is a pseudoinverse of \(f\), for all \(a\in A\).
\end{lemma}

\begin{proof}
We define an independent random variable \(\A\sim A:=X^Y\) as a sequence of independent random variables \(\A=(\A_y)_{y\in Y}\) where each \(\A_y\sim X\) is defined as
\[
\Pr[\A_y=x]:=\Pr[\X=x\mid f(\X)=y]\text{ for all }x\in X.
\]
If we then define the deterministic function \(g:Y\times A\to V\) as
\[
g(y,a):=a_y,
\]
it directly follows that \(g(f(\X),\A)\equiv\X\) via the law of total probability, as desired.
\end{proof}

\subsection{Extractors}

This section collects additional basic definitions of randomness extractors and useful auxiliary results.

\begin{definition}[Strong seeded extractor]
    A function $\Ext:\F_2^n\times\F_2^s\to\F_2^m$ is said to be a \emph{$(k,\eps)$-strong seeded extractor} if for every $(n,k)$-source $\X$ it holds that
    \begin{equation*}
        \Ext(\X,\U_s), \U_s\approx_\eps \U_m,\U_s,
    \end{equation*}
    where $\U_s$ is independent of $\X$ and $\U_m$.
    Moreover, we say that $\Ext$ is \emph{linear} if $\Ext(\cdot,y)$ is a linear function for all $y\in\F_2^s$.
\end{definition}

\begin{lemma}[Leftover Hash Lemma~\cite{HILL99}]\label{lem:lhl}
For every $0<k<n$, $\eps>0$, and $m\leq k-2\log(1/\eps)$ there exists an explicit linear $(k,\eps)$-strong seeded extractor $\Ext:\F_2^n\times\F_2^n\to\F_2^m$.
\end{lemma}

We will use the following simple corollary of \cref{lem:lhl}.
\begin{lemma}\label{lem:lhlsurj}
    For every $n>k>100$ there exists a family of linear maps $L:\F_2^n\to\F_2^{k/100}$ such that for every $(n,k)$-source $\X$ we have that at least a $0.8$-fraction of linear maps in the family are surjective on $\supp(\X)$.
\end{lemma}
\begin{proof}
    We apply the leftover hash lemma with $\eps=2^{-k/5}$ and $m=k/100\leq k-2\log(1/\eps)=3k/5$.
    Fix any flat $(n,k)$-source $\X$.
    By an averaging argument, we have $L(\X)\approx_{10\eps} \U_1$ for more than a $0.8$-fraction of linear maps $L$ in the family given by the leftover hash lemma.
    Since $10\eps<2^{-k/100}=2^{-m}$ when $k>100$, we conclude that every such map $L$ is surjective on $\supp(\X)$.
\end{proof}

The lemma below is a special case of a much more general result of Dhar and Dvir on leftover hashing with $\ell_\infty$ guarantees~\cite[Theorem II.4]{DD22}.

\begin{lemma}[\protect{\cite[Theorem II.4, special case]{DD22}}]\label{lem:ellInftyLHL-new}
There exists a constant \(C>0\) such that for all \(n\geq k\geq2\), the following holds. Fix any set \(S\subseteq\F_2^n\) of size \(2^k\). Then at least \(0.99\) fraction all linear maps \(L:\F_2^n\to\F_2^m\) with output length \(m=\lfloor k-C\log k\rfloor\) are surjective when restricted to \(S\).
\end{lemma}

We now define various special types of extractors and dispersers.

\begin{definition}[Two-source disperser]
    A function $\Disp:\F_2^n\times\F_2^n\to\F_2$ is said to be a \emph{$k$-two-source disperser} if for any two independent $(n,k)$-sources $\X$ and $\Y$ it holds that $\supp(\Disp(\X,\Y))=\zo$.
\end{definition}

\begin{definition}[Two-source extractor]
    A function $\Ext:\F_2^n\times\F_2^n\to\F_2^m$ is said to be a \emph{$(k,\eps)$-two-source extractor} if for any two independent $(n,k)$-sources $\X$ and $\Y$ it holds that
    \begin{equation*}
        \Ext(\X,\Y)\approx_\eps \U_m.
    \end{equation*}
\end{definition}

\begin{definition}[Sumset disperser]\label{def:sumsetdisp}
    A function $\Disp:\F_2^n\to\F_2$ is said to be a \emph{$k$-sumset disperser} if for any $(n,k)$-sumset source $\W$ it holds that $\supp(\Disp(\W))=\zo$.
\end{definition}

\begin{definition}[Sumset extractor]
    A function $\Ext:\F_2^n\to\F_2^m$ is said to be a \emph{$(k,\eps)$-sumset extractor} if for any $(n,k)$-sumset source $\W$ it holds that
    \begin{equation*}
        \Ext(\W)\approx_\eps \U_m.
    \end{equation*}
\end{definition}

Note that every $k$-sumset disperser $\Disp$ induces a $k$-two-source disperser $\Disp'$ by setting $\Disp'(x,y)=\Disp(x+y)$.
The same holds in the setting of extractors.

\dobib

\section{Low-degree polynomials extract from small families}\label{sec:smallfamilies}

\subsection{Low-degree polynomials extract from a single source}\label{sec:biasweak}

In this section we prove \cref{our-results:main:first-theorem}, which we restate here.

\weaksource*

Recall that for a function $f:\F_2^n\to\F_2$ and a random variable $\X\sim \F_2^n$ we define the bias of $f$ with respect to $\X$ as
\begin{equation*}
    \bias_{\X}(f) = \Pr_{x\sim \X}[f(x)=0]-\Pr_{x\sim \X}[f(x)=1]=\E_{x\sim \X}\left[(-1)^{f(x)}\right].
\end{equation*}
When $\X$ is uniformly distributed over a set $X\subseteq\F_2^n$, we may write $\bias_{\X}(f)=\bias_{X}(f)$, and when $X=\F_2^n$ we simply write $\bias(f)$ for $\bias_{X}(f)$. 
We will need the following characterization about the moments of the bias, whose statement and proof are analogous to those of~\cite[Claim 2.1]{BHL12} for the special case of a uniform input distribution.
\begin{restatable}[\protect{Simple extension of \cite[Claim 2.1]{BHL12}}]{lemma}{biasmoment}
\label{lem:biasmoment}
    If $f:\F_2^n\to\F_2$ is a random degree $d$ polynomial, it holds that
    \begin{equation*}
        \E_f\left[\bias_{\X}(f)^t\right]=\Pr_{x^{(1)},\dots,x^{(t)}\sim\X}[\forall p:\F_2^n\to\F_2,\deg(p)\leq d: p(x^{(1)})+\cdots+ p(x^{(t)})=0]
    \end{equation*}
    for any random variable $\X\sim\F_2^n$.
\end{restatable}
\begin{proof}
    The argument follows the same lines as that behind~\cite[Claim 2.1]{BHL12}.
    Observe that the lemma statement follows immediately if we show that
    \begin{equation*}
        \E_f[\bias_{\X}(f)^t] = \Pr_{x^{(1)},\dots,x^{(t)}\sim \X}\left[\eval_d(x^{(1)})+\cdots+\eval_d(x^{(t)})=0\right].
    \end{equation*}
    
    We begin by writing $f$ as
    \begin{equation*}
        f(x) = \sum_{I\subseteq[n]:|I|\leq d} \alpha_I x_I,
    \end{equation*}
    where the $\alpha_I$ are independent and uniformly distributed over $\F_2$ and $x_I=\prod_{i\in I}x_i$.
    Then, we have that
    \begin{align*}
        \E_f[\bias_{\X}(f)^t] &= \E_{\{\alpha_I\}}\left[\prod_{j=1}^t\E_{x^{(j)}\sim\X}\left[(-1)^{\sum_{I\subseteq[n]:|I|\leq d} \alpha_I x^{(j)}_I}\right]\right]\\
        & = \E_{\{\alpha_I\}}\left[\E_{x^{(1)},\dots,x^{(t)}\sim\X}\left[\prod_{j=1}^t(-1)^{\sum_{I\subseteq[n]:|I|\leq d} \alpha_I x^{(j)}_I}\right]\right]\\
        &= \E_{x^{(1)},\dots,x^{(t)}\sim\X}\left[\E_{\{\alpha_I\}}\left[(-1)^{\sum_{I\subseteq[n]:|I|\leq d} \alpha_I \left(\sum_{j=1}^t x^{(j)}_I\right)}\right]\right]\\
        &= \E_{x^{(1)},\dots,x^{(t)}\sim\X}\left[\prod_{I\subseteq[n]:|I|\leq d} \E_{\alpha_I}\left[(-1)^{\alpha_I \left(\sum_{j=1}^t x^{(j)}_I\right)}\right]\right]\\
        &= \E_{x^{(1)},\dots,x^{(t)}\sim\X}\left[\prod_{I\subseteq[n]:|I|\leq d} \mathbf{1}_{\left\{x^{(1)}_I+\cdots+x^{(t)}_I=0\right\}}\right]\\
        &= \E_{x^{(1)},\dots,x^{(t)}\sim\X}\left[\mathbf{1}_{\left\{\eval_d(x^{(1)})+\cdots+\eval_d(x^{(t)})=0\right\}}\right]\\
        &= \Pr_{x^{(1)},\dots,x^{(t)}\sim\X}\left[\eval_d(x^{(1)})+\cdots+\eval_d(x^{(t)})=0\right],
    \end{align*}
    as desired.
\end{proof}

We are now ready to prove \cref{our-results:main:first-theorem}.
\begin{proof}[Proof of \cref{our-results:main:first-theorem}]
    The proof follows along the same lines as the proof of~\cite[Lemma 1.2]{BHL12}, combined with a linear hashing trick.
    For an integer $t>0$, we focus on bounding the $t$-th moment of $\bias_{\X}(f)$.
    By \cref{lem:biasmoment}, we have that
    \begin{equation*}
        \E_f\left[\bias_{\X}(f)^t\right] = \Pr_{x^{(1)},\dots,x^{(t)}\sim \X}[\forall p:\F_2^n\to\F_2, \deg(p)\leq d: p(x^{(1)})+\cdots+p(x^{(t)})=0].
    \end{equation*}
    Fix any linear map $L:\F_2^n\to\F_2^m$.
    Observe that
    \begin{multline}\label{eq:UBlinmap}
        \Pr_{x^{(1)},\dots,x^{(t)}\sim \X}[\forall p:\F_2^n\to\F_2, \deg(p)\leq d: p(x^{(1)})+\cdots+p(x^{(t)})=0]\\
        \leq \Pr_{x^{(1)},\dots,x^{(t)}\sim \X}[\forall p:\F_2^m\to\F_2, \deg(p)\leq d: p(L(x^{(1)}))+\cdots+p(L(x^{(t)}))=0].
    \end{multline}
    This holds since $\deg(p\circ L)\leq \deg(p)\leq d$, where $p\circ L$ denotes the composition of the polynomial $p$ and the linear map $L$.
    Then, \cref{lem:biasmoment} yields
    \begin{align*}
        &\Pr_{x^{(1)},\dots,x^{(t)}\sim \X}[\forall p:\F_2^m\to\F_2, \deg(p)\leq d: p(L(x^{(1)}))+\cdots+p(L(x^{(t)}))=0]\\
        &= \Pr_{w^{(1)},\dots,w^{(t)}\sim L(\X)}[\forall p:\F_2^m\to\F_2, \deg(p)\leq d: p(w^{(1)})+\cdots+p(w^{(t)})=0]\\
        &= \E_g\left[\bias_{L(\X)}(g)^t\right],
    \end{align*}
    where the expectation is taken over the choice of a random degree $d$ polynomial $g:\F_2^m\to\F_2$.
    Therefore, we conclude that
    \begin{equation}\label{eq:biasUBLinMaps}
        \E_f\left[\bias_{\X}(f)^t\right]\leq \E_g\left[\bias_{L(\X)}(g)^t\right]
    \end{equation}
    for all linear maps $L$.

    Let $c>0$ be the absolute constant from \cref{intro:thm:BHL12}.
    Without loss of generality, we enforce that $c<\min(1/4,\delta/2)$ (if this does not hold for the choice of $c$ from~\cref{intro:thm:BHL12}, take a smaller $c$).
    Since $\minH(\X)=k$, the leftover hash lemma (\cref{lem:lhl}) guarantees the existence of a linear map $L:\F_2^n\to\F_2^{m}$ with $m=k(1-2c/d)$ such that
    \begin{equation}\label{eq:SDlinmap}
        L(\X)\approx_{2^{-ck/d}} \U_{m}.
    \end{equation}
    Note that, by our choice of $c$, we have that $m\geq (1-\delta)k\geq d$.
    \cref{eq:SDlinmap} implies that
    \begin{equation}\label{eq:linbiasbeforesimpli}
        \Pr_g\left[|\bias_{L(\X)}(g)|>2^{-\frac{ck}{2d}+1}\right]\leq \Pr_g\left[|\bias(g)|>2^{-\frac{ck}{2d}+1}-2^{-\frac{ck}{d}}\right] \leq 2^{-c\binom{m}{\leq d}}.
    \end{equation}
    The last inequality follows from \cref{intro:thm:BHL12} applied to $\F_2^m$ because
    \begin{equation*}
        2^{-\frac{ck}{2d}+1}-2^{-\frac{ck}{d}} \geq 2^{-\frac{ck}{2d}} \geq 2^{-cm/d},
    \end{equation*}
    since $m=k(1-2c/d)\geq k/2$ as we enforced that $c<1/4$.
    Since
    \begin{equation*}
        \binom{m}{\leq d}=\binom{k(1-2c/d)}{\leq d}\geq \alpha \binom{k}{\leq d}
    \end{equation*}
    for some constant $\alpha>0$ depending only on $c$ and $\delta$, we get from \cref{eq:linbiasbeforesimpli} that
    \begin{equation}\label{eq:linbiasaftersimpli}
        \Pr_g\left[|\bias_{L(\X)}(g)|>2^{-c_1k/d}\right]\leq 2^{-c_1\binom{k}{\leq d}}
    \end{equation}
    for some absolute constant $c_1>0$.

    We now proceed to bound the $t$-th moment $\E_f\left[\bias_{\X}(f)^t\right]$ for an appropriate $t$.
    Set $t=c_2\cdot\frac{d}{k}\cdot \binom{k}{\leq d}$ for a sufficiently large constant $c_2>0$ depending only on $c_1$.
    Using \cref{eq:biasUBLinMaps,eq:linbiasaftersimpli}, we have that
    \begin{align*}
        \E_f\left[\bias_{\X}(f)^t\right]&\leq \E_g\left[\bias_{L(\X)}(g)^t\right]\\
        &\leq 2^{-c_1kt/d}+2^{-c_1\binom{k}{\leq d}}\\
        &\leq 2^{-c_3 \binom{k}{\leq d}},
    \end{align*}
    where the last equality uses our choice of $t$ and holds for a sufficiently small constant $c_3>0$.
    Let $c_4 = c_3/2$.
    Combining the bound above with Markov's inequality, we conclude that
    \begin{equation*}
        \Pr_f\left[|\bias_{\X}(f)|>2^{-c_4 k/d}\right]\leq \frac{\E_f\left[\bias_{\X}(f)^t\right]}{2^{-c_4 tk/d}} \leq 2^{-c_4 \binom{k}{\leq d}},
    \end{equation*}
    which yields the desired lemma statement with absolute constant $c_4>0$.
\end{proof}

\subsection{Low-degree polynomials extract from small families}\label{sec:applications}

We showcase applications of \cref{our-results:main:first-theorem} to some important small families of sources.

\subsubsection*{Local sources}

First, we consider the scenario of locally-samplable sources~\cite{DW12,Vio14,ACGLR22}.
A source $\X\sim\F_2^n$ is said to be \emph{$r$-local} if $\X = g(\U_m)$, where $g:\F_2^m\to\F_2^n$ is some function such that each output bit depends on at most $r$ input bits and $\U_m$ denotes the uniform distribution over $\F_2^m$.
We can take $m\leq nr$ without loss of generality, in which case there are at most
\begin{equation*}
    \left(\binom{nr}{r}2^{2^r}\right)^n \leq (en)^{r n}\cdot 2^{2^r\cdot n} \leq 2^{2r n\log n}\cdot 2^{2^r\cdot n} = 2^{n(2r \log n+2^r)}
\end{equation*}
$r$-local sources of length $n$, provided that $n\geq 4$.
If $k=Cd n^{1/d}(r\log n+2^r)^{1/d}$, we have that
\begin{equation*}
    \binom{k}{\leq d}\geq \binom{k}{d} \geq (k/d)^d = C^d n(r \log n+2^r).
\end{equation*}
Let $c$ be the absolute constant from \cref{our-results:main:first-theorem}.
If we take $C$ to be large enough depending only on $c$ so that $c\cdot C^d n(r \log n+2^r) > n(2r \log n+2^r)$, we conclude that there are fewer than $2^{c\binom{k}{\leq d}}$ sources in the family.
Combining a union bound over these sources with \cref{our-results:main:first-theorem} immediately yields the following result.

\begin{corollary}[Low-degree polynomials are good local source extractors]\label{coro:improveLocal}
    For every $\delta\in(0,1)$ there exist constants $c,C>0$ such that the following holds for all $n$: For $k\geq Cd n^{1/d}(r\log n+2^r)^{1/d}$ and $d\leq (1-\delta)k$, a random degree $d$ polynomial is an $(\eps=2^{-ck/d})$-extractor for the family of length-$n$ $r$-local sources of min-entropy $k$ with probability at least $1-2^{-c\binom{k}{\leq d}}$ over the choice of $f$.
\end{corollary}

\cref{coro:improveLocal} both improves on and simplifies the proof of~\cite[Theorem 1.1]{ACGLR22}, which required min-entropy $k\geq C 2^r r^2 d(2^r n\log n)^{1/d}$ and error $\eps = 2^{-\frac{ck}{d^3 2^r r^2}}$ for some absolute constants $c,C>0$ and an intricate initial reduction to local non-oblivious bit-fixing sources.
In particular, observe that when $d>r$, we now get that the min-entropy is $O(d(n\log n)^{1/d})$.
Previously, this only happened when $d>2^r$.

It is also instructive to compare our improved result with the lower bound from~\cite[Theorem 1.2]{ACGLR22}, which states that no degree-$d$ polynomial extracts from length-$n$ $r$-local sources of min-entropy $k=cd(r n\log n)^{1/d}$ for some absolute constant $c>0$.
For example, when the locality satisfies $r<\log\log n$, \cref{coro:improveLocal} is optimal up to the constant factor.

\subsubsection*{Polynomial sources}
A random variable $\X\sim\F_2^n$ is a degree-$r$ polynomial source if there exist $\F_2$-polynomials $p_1,\dots,p_n:\F_2^m\to\F_2$ of degree at most $r$ for some positive integer $m$ (the \emph{input length}) such that $\X = P(\U_m)$, where $P=(p_1,\dots,p_n)$ and $\U_m$ is uniform over $\F_2^m$. 
This is a very challenging model to extract from, and several papers have attempted to do so \cite{DGW09,BG13,GVJZ23,CGG24}. 
In the most recent work, Chattopadhyay, Goodman, and Gurumukhani~\cite{CGG24} established the following input reduction lemma.
\begin{lemma}[Input reduction lemma for polynomial sources \protect{\cite[Lemma 1]{CGG24}}]\label{lem:input-red-poly-sources}
    There exists a constant $\alpha>0$ such that the following holds for all positive integers $n, k,r$ such that $k\leq n$: Suppose that $\X\sim\F_2^n$ is a degree-$r$ polynomial source with min-entropy $k$.
    Then, $\X\approx_{2^{-k}} \Y$, where $\Y\sim\F_2^n$ is a convex combination of degree-$r$ polynomial sources with min-entropy $k'=k-1$ and input length $m=\alpha (k-1)$.
\end{lemma}

Given \cref{lem:input-red-poly-sources}, in order to show that a function $f:\F_2^n\to\F_2$ is an $(\eps+2^{-k})$-extractor for degree-$r$ polynomial sources with min-entropy $k$ it suffices to show that $f$ is an $\eps$-extractor for degree-$r$ polynomial sources with min-entropy $k-1$ and input length $m=\alpha (k-1)$.
There are at most
\begin{equation*}
    2^{\binom{m}{\leq r}n} \leq 2^{\left(\frac{e m}{r}\right)^r n} =2^{\left(\frac{\beta (k-1)}{r}\right)^r n}
\end{equation*}
such sources, where $\beta=e\cdot \alpha$ and we have used the inequality $\binom{m}{\leq r}\leq \left(\frac{e\cdot m}{r}\right)^r$ valid for all $m\geq 1$ and $1\leq r\leq m$.
Set $k= C\left(\frac{C^r d^{d}n}{r^{r}}\right)^{\frac{1}{d-r}}$.
Then, if we take $C$ to be a sufficiently large constant depending on $\beta$ and the constant $c>0$ guaranteed by \cref{our-results:main:first-theorem}, we get that
\begin{equation*}
    2^{\left(\frac{\beta (k-1)}{r}\right)^r n} < 2^{c\left(\frac{k-1}{d}\right)^d}\leq 2^{c\binom{k-1}{\leq d}}.
\end{equation*}
Therefore, we can apply \cref{our-results:main:first-theorem} to the family of degree-$r$ polynomial sources with min-entropy $k-1$ and input length $m=\alpha (k-1)$ to obtain the following result.

\begin{corollary}[Low-degree polynomials extract from polynomial sources]\label{coro:lowDegPolySources}
    For every $\delta\in(0,1)$ there exist constants $c,C>0$ such that the following holds for all integers $n\geq 2$: 
    For $k\geq 2$ satisfying $k\geq C\left(\frac{C^r d^{d}n}{r^{r}}\right)^{\frac{1}{d-r}}$ and $d\leq (1-\delta)k$, a random polynomial of degree at most $d$ is an $(\eps=2^{-ck/d})$-extractor for the family of length-$n$ degree-$r$ polynomial sources of min-entropy $k$ with probability at least $1-2^{-c\binom{k}{\leq d}}$ over the choice of $f$.

    In particular, if we take $d=2r$ then the above holds for any $k\geq C d n^{2/d}$ for a sufficiently large constant $C>0$.
\end{corollary}

\subsubsection*{Variety sources}
A random variable $\X\sim\F_2^n$ is a variety source of degree $r$ and min-entropy $k$ if for some $t$ there exist polynomials $p_1,\dots,p_t:\F_2^n\to\F_2$ of degree at most $r$ such that $\X$ is uniformly distributed over the variety
\begin{equation*}
    V(p_1,\dots,p_t) =\{x\in\F_2^n\mid p_1(x)=p_2(x)=\cdots=p_t(x)=0\}.
\end{equation*}
This family of sources has received significant interest over the last decade, both over $\F_2$ and larger fields~\cite{Dvi09,Rem16,LZ19,GVJZ23}.
Sufficiently strong explicit extractors for variety sources are known to imply breakthrough circuit lower bounds~\cite{GKW21}. In order to bound the number of variety sources we need to handle, we use the following structural result implicit in~\cite[Theorem 5.1]{CT15}, which states that we can assume the variety is defined by $\leq n+1$ polynomials.
We provide the short proof for completeness.
\begin{lemma}[Input reduction lemma for variety sources \protect{\cite[Theorem 5.1]{CT15}}]\label{lem:inputRedVariety}
    For any variety $V(p_1,\dots,p_t)$ of degree $r$, there exists a variety $V(q_1,\dots,q_\ell)$ of degree $r$ such that $V(p_1,\dots,p_t)=V(q_1,\dots,q_\ell)$ and $\ell= n+1$.
\end{lemma}
\begin{proof}
    To define the polynomials $q_i$, first suppose that we sample coefficients $\alpha_{ij}$ uniformly at random from $\F_2$, and set $q_i(x) =\sum_{j=1}^t \alpha_{ij} p_j(x)$.
    Then, all the $q_i$'s have degree at most $r$ and $V(p_1,\dots,p_t)\subseteq V(q_1,\dots,q_\ell)$.
    Moreover, for each $x\not\in V(p_1,\dots,p_t)$ we have that $\Pr[q_i(x)=0\text{ for all }i\in[\ell]]=2^{-\ell}=2^{-(n+1)}$.
    This means the expected number of $x$'s such that $x\in V(q_1,\dots,q_\ell)\setminus V(p_1,\dots,p_t)$ is at most $2^n\cdot 2^{-(n+1)}=1/2$. Thus, there is a fixing of the coefficients $(\alpha_{ij})$ such that $V(q_1,\dots,q_\ell)=V(p_1,\dots,p_t)$.
\end{proof}

With the help of \cref{lem:inputRedVariety}, we conclude that the number of variety sources of degree $r$ and min-entropy $k$ is at most $2^{(n+1)\binom{n}{\leq r}}\leq 2^{(n+1)^{r+1}}$.
If $k=Cd (n+1)^{\frac{r+1}{d}}$, we have that
\begin{equation*}
    \binom{k}{\leq d}\geq \binom{k}{d}\geq (k/d)^d = C^d (n+1)^{r+1}.
\end{equation*}
Let $c>0$ be the absolute constant from \cref{our-results:main:first-theorem}.
Then, if we take $C>0$ to be sufficiently large depending only on $c$ so that $c\cdot C^d n^{r+1}> (n+1)^{r+1}$, we get that the family of variety sources of degree $r$ and min-entropy $k$ has size smaller than $2^{c\binom{k}{\leq d}}$, and so combining \cref{our-results:main:first-theorem} and a union bound immediately yields the following result.

\begin{corollary}[Low-degree polynomials extract from variety sources]\label{coro:lowDegVarietySources}
    For every $\delta\in(0,1)$ there exist constants $c,C>0$ such that the following holds for all $n$: For $k\geq Cd n^{\frac{r+1}{d}}$ and $d\leq (1-\delta)k$, a random polynomial of degree $\leq d$ is an $(\eps=2^{-ck/d})$-extractor for the family of length-$n$ variety sources of degree $r$ and min-entropy $k$ with probability at least $1-2^{-c\binom{k}{\leq d}}$ over the choice of $f$.
\end{corollary}

The best explicit extractors for variety sources over $\F_2$ either work for constant degree $r$ and min-entropy $k=(1-c_r)n$~\cite{LZ19}, or large degree $r=n^\alpha$ and very high min-entropy $n-n^\beta$ with $\alpha+\beta<1/2$~\cite{Rem16}.
By~\cite{GKW21}, an explicit version of \cref{coro:lowDegVarietySources} would be more than enough to imply significantly improved circuit lower bounds.

\subsection{Low-degree polynomials are good linear seeded extractors}\label{sec:seeded}

In this section we study the parameters of strong linear seeded extractors computable by low-degree polynomials. Such objects are crucial in the construction of several affine extractors \cite{Li11,li2023explicit}. We start with the following basic definitions.

\begin{definition}[Almost balanced codes]
We say that a string \(x\in\F_2^n\) is \(\eps\)-balanced if it has Hamming weight \(\frac{1-\eps}{2}n\leq\mathsf{wt}(x)\leq\frac{1+\eps}{2}n\). 
We say that a code \(C\subseteq\F_2^n\) is \(\eps\)-balanced\footnote{Another standard name for this notion is ``$\eps$-biased code.''} if all nonzero \(x\in C\) are \(\eps\)-balanced. We say that a code \(C\subseteq\F_2^n\) is \(\delta\)-almost \(\eps\)-balanced if \(\Pr_{x\sim C}[x\text{ is not \(\eps\)-balanced}]\leq\delta\).
\end{definition}

\begin{definition}[Reed-Muller codes]
    We define the Reed-Muller code $\mathsf{RM}(n,d)$ as
    \begin{equation*}
        \mathsf{RM}(n,d) = \{(f(x))_{x\in\F_2^n}\mid f:\F_2^n\to\F_2, \deg f\leq d\}.
    \end{equation*}
\end{definition}

We will need to import the following results. The first corollary about balancedness of $\mathsf{RM}(n,d)$ follows directly from~\cite[Lemma 1.2]{BHL12} (\cref{intro:thm:BHL12}), which corresponds to setting $k=n$ in \cref{our-results:main:first-theorem}.

\begin{corollary}\label{cor:almost-balanced-RM-code}
    For every $\gamma\in (0,1)$ there exists a constant $c>0$ such that the Reed-Muller code \(\mathsf{RM}(n,d)\) with $d\leq (1-\gamma)n$ is \(\delta\)-almost \(\eps\)-balanced, where \(\delta=2^{-c\binom{n}{\leq d}}\) and \(\eps=2^{-cn/d}\).
\end{corollary}

\begin{lemma}[\protect{Johnson bound~\cite[Section 7.3]{GRS22}, adapted}]\label{lem:johnson-bound}
Every linear \(\eps\)-balanced code of length $T$ has list-size at most \(2T\) at radius \(\frac{1-\sqrt{\eps}}{2}\).\footnote{A code $C\subseteq \F_q^n$ has list-size at most $L$ at radius $\rho$ if $|\{c\in C:\mathsf{wt}(c-x)\leq \rho n\}|\leq L$ for all vectors $x\in\F_q^n$.}
\end{lemma}

The following well-known lemma connects the list-decodability of a linear code and its performance as a seeded extractor.
See \cite{shaltiel2011introduction} for a nice exposition of this result.
\begin{lemma}[Extractors from codes \cite{trevisan2001extractors}]\label{lem:code-to-extractor}
Let \(G\in\F_2^{T\times n}\) be the generator matrix of a linear code with list size \(\leq L\) at radius \(\leq\frac{1-\eps}{2}\), and let $t=\log T$. 
Then the function \(\Ext_G:\F_2^n\times\F_2^t\to\F_2\) defined as \(\Ext_G(x,y):=(Gx)_y=\langle x,G_y\rangle\) is a \((k,\eps)\)-strong seeded extractor for min-entropy \(k=\log L + \log(1/\eps) + 1\).
\end{lemma}

Using these, we are ready to show that low-degree polynomials can function as strong linear seeded extractors. Our key tool is the following.

\begin{lemma}[Random subcodes of almost-balanced codes are balanced]\label{lem:random-subcodes-are-balanced}
    Let \(G\in\F_2^{n\times k}\) be the generator matrix of a \(\delta\)-almost \(\eps\)-balanced code. Then, for a uniformly random \(H\in\F_2^{k\times\ell}\) we have that
    \[
    \Pr_H[GH\text{ is not the generator matrix of an \(\eps\)-balanced code}]\leq\delta\cdot(2^\ell-1).
    \]
\end{lemma}
\begin{proof}
When we refer to a matrix \(G\) as a code, we are referring to the code generated by \(G\). We have
\begin{align*}
\Pr_H[GH\text{ is not an \(\eps\)-balanced code}]&=\Pr_H[\exists x\neq 0 : (GH)x\text{ is not \(\eps\)-balanced}]\\
&=\Pr_H\left[\bigvee_{x\neq 0}(GH)x\text{ is not \(\eps\)-balanced}\right]\\
&\leq\sum_{x\neq 0}\Pr_H[(GH)x\text{ is not \(\eps\)-balanced}]\\
&=\sum_{x\neq 0}\Pr_{y\sim\F_2^k}[Gy\text{ is not \(\eps\)-balanced}]\\
&\leq(2^\ell-1)\cdot\delta.\qedhere
\end{align*}
\end{proof}

\begin{theorem}[Strong seeded extractors from random subcodes of linear almost-balanced codes]\label{thm:general-but-somewhat-intermediate-thm}

Let \(G\in\F_2^{T\times r}\) be the generator matrix of a \(\delta\)-almost \(\eps\)-balanced code and let $t=\log T$. 
For any matrix \(H\in\F_2^{r\times n}\), define the function \(\Ext_{GH}:\F_2^n\times\F_2^t\to\F_2\) as \(\Ext_{GH}(x,y):=(GHx)_y\). Then for \(k:=\log(2T)+\log(1/\sqrt{\eps})+1\) and a uniformly random \(H\), it holds that
\[
\Pr_H[\Ext_{GH}\text{ is not a \((k,\sqrt{\eps})\)-strong seeded extractor}]\leq\delta\cdot2^n.
\]
\end{theorem}
\begin{proof}
By \cref{lem:random-subcodes-are-balanced}, we know that \(GH\) is the generator of an \(\eps\)-balanced code, except with probability at most \(\delta\cdot2^n\). By the Johnson bound (\cref{lem:johnson-bound}), this implies that \(GH\) is the generator of a code with list size at most \(2T\) at radius at most \(\frac{1-\sqrt{\eps}}{2}\), except with probability at most \(\delta\cdot2^n\). This immediately implies via \cref{lem:code-to-extractor} that \(\Ext_{GH}\) is a \((k,\sqrt{\eps})\)-strong seeded extractor, except with probability at most \(\delta\cdot2^n\).
\end{proof}
Using this, we obtain the following.
\begin{theorem}[Strong seeded extractors from random subcodes of Reed-Muller codes]\label{thm:highly-general-reed-muller}
For any \(\gamma>0\) there exists a constant \(c=c(\gamma)>0\) such that for any \(d\leq(1-\gamma)n\), the following holds. Let \(G\in\F_2^{T\times\binom{t}{\leq d}}\) be a generator matrix of the Reed-Muller code \(\mathsf{RM}(t,d)\). For any matrix \(H\in\F_2^{\binom{t}{\leq d}\times n}\), define the function \(\Ext_{GH}:\F_2^n\times\F_2^t\to\F_2\) as \(\Ext_{GH}(x,y):=(GHx)_y\). Then for \(k:=\log(2T)+\log(1/\sqrt{\eps})+1\) and \(\eps:=2^{-ct/d}\) and \(\delta:=2^{-c\binom{t}{\leq d}}\) and a uniformly random \(H\), it holds that
\[
        \Pr_H\left[\Ext_{GH}\text{ is not a \((k,\sqrt{\eps})\)-strong seeded extractor}\right]\leq\delta\cdot2^n.
\]
    Furthermore, for any \(H\) it holds that \(\Ext_{GH}\) has left-degree at most \(1\) and right-degree at most \(d\), i.e., any monomial in \(\Ext_{GH}\) contains at most one variable from the left source and at most $d$ variables from the right source.
\end{theorem}
\begin{proof}
The claim that \(\Ext_{GH}\) is a strong seeded extractor follows immediately by combining the fact that the Reed-Muller code \(\mathsf{RM}(t,d)\) is linear and \(\delta\)-almost \(\eps\)-balanced (\cref{cor:almost-balanced-RM-code}) with the fact that random subcodes of linear \(\delta\)-almost \(\eps\)-balanced codes give strong seeded extractors (\cref{thm:general-but-somewhat-intermediate-thm}). All that remains is to show that \(\Ext_{GH}\) will have left-degree at most \(1\) and right-degree at most \(d\) for any \(H\). To see why, recall that, since \(G\in\F_2^{T\times\binom{t}{\leq d}}\) is a generator of \(\mathsf{RM}(t,d)\), there exist degree \(\leq d\) polynomials \(p_1,\dots,p_{\binom{t}{\leq d}}:\F_2^t\to\F_2\) whose truth tables make up the columns of \(G\). Thus, if we let \(L_i:\F_2^n\to\F_2\) define the linear function \(L_i(x):=\langle H_i,x\rangle\) for each row \(H_i\) of \(H\), we can write \(\Ext(x,y)\) as
\[
\Ext_{GH}(x,y):=(GHx)_y=\sum_{i\in\binom{t}{\leq d}}p_i(y)\cdot L_i(x),
\]
and thus it has left-degree at most \(1\) and right-degree at most \(d\), as desired.
\end{proof}

This immediately gives the following.

\begin{theorem}\label{thm:nice-parameter-setting-sExt}
For any \(\gamma\in(0,1)\) there exist constants \(c,C>0\) such that the following holds. For any \(d\leq(1-\gamma)n\) and \(k\geq Cdn^{1/(d-1)}\), there exists a linear \((k,\eps)\)-strong seeded extractor \(\Ext:\F_2^n\times\F_2^k\to\F_2\) of degree \(\leq d\) with error \(\eps=2^{-ck/d}\).
\end{theorem}
\begin{proof}
Set \(t=k=O(dn^{1/(d-1)})\) in \cref{thm:highly-general-reed-muller}.
\end{proof}

\dobib

\section{Low-degree polynomials extract from sumset sources}\label{sec:largefamilies}

\subsection{Low-degree polynomials disperse from sumset sources}\label{sec:lowdegsumsetdisp}

As a warmup, we prove a disperser version of \cref{thm:lowdegsumsetext} (with a slightly better bound on the probability). For simplicity, we focus here on the case of even degree \(d\).

\begin{theorem}\label{thm:lowdegsumsetdisp}
    There exist constants $C,c>0$ such that for any \(n\geq k\geq d\in\N\) (with \(d\) even) satisfying \(d\leq\frac{c\log n}{\log\log n}\) and $k\geq C dn^{2/d}$, a random degree $d$ polynomial $f:\F_2^n\to\F_2$ is a disperser for \((n,k)\)-sumset sources, with probability at least $1-2^{-c\binom{k}{\leq d}}$.
\end{theorem}

Our key lemma (recall the informal \cref{claim:key1}) states that for any two large subsets $A$ and $B$ we can find appropriately small subsets $A'\subseteq A$ and $B'\subseteq B$ such that $\rank(\eval_d(A'+B'))$ is large.

\begin{restatable}{lemma}{subsetrank}\label{lem:subsetrank}
    There exists a constant $C>0$ such that for all \(n\geq k\geq d\in\N\) (with \(d\) even) satisfying \(k\geq C(1+\log n)\), the following holds. Let $A,B\subseteq\F_2^n$ be sets of size $2^k$.
    Then, there exist subsets $A'\subseteq A$ and $B'\subseteq B$ such that $|A'|,|B'|= \binom{k - C\log n}{\leq d/2}$ and $\rank(\eval_d(A'+B'))\geq \binom{k-C\log n}{\leq d}$.
\end{restatable}

Before we proceed to the (simple) proof of \cref{lem:subsetrank}, we note some basic properties of evaluation vectors.
First, it is well known that $\rank(\eval_d(\cB^n_d(0))) = \binom{n}{\leq d}$, as $\cB^n_d(0)$ is an interpolating set for degree $\leq d$ polynomials (see, e.g., \cite[Proposition 6.21]{o2014analysis}).
Moreover, we have the following property.

\begin{claim}\label{lem:ranklinmap}
If $L:\F_2^n\to\F_2^m$ is a linear map, then
\begin{equation*}
    \rank(\eval_d(S))\geq \rank(\eval_d(L(S)))
\end{equation*}
for every set $S\subseteq\F_2^n$.
\end{claim}
\begin{proof}
    Suppose that $x_1,\dots,x_r\in S$ are distinct vectors such that the associated evaluation vectors $\eval_d(x_1),\dots,\eval_d(x_r)$ satisfy $\sum_{i=1}^r \eval_d(x_i)=0$.
    Equivalently, for every polynomial $f$ of degree at most $d$ it holds that
    \begin{equation*}
        \sum_{i=1}^r f(x_i)=0.
    \end{equation*}
    
    It suffices to show that the vectors $\eval_d(L(x_1)),\dots,\eval_d(L(x_r))$ also satisfy
    \begin{equation}\label{eq:lindeplinmap}
        \sum_{i=1}^r \eval_d(L(x_i))=0.
    \end{equation}
    Fix a set $U\subseteq [m]$ such that $|U|\leq d$.
    Note that $L(x)^U=\prod_{j\in U} L(x)_j$ can be written as $f_{L,U}(x)$ for some polynomial $f_{L,U}$ of degree at most $|U|\leq d$ which depends only on $L$ and $U$.
    Therefore, we have
    \begin{equation*}
        \sum_{i=1}^r L(x_i)^U = \sum_{i=1}^r f_{L,U}(x_i)=0,
    \end{equation*}
    which establishes \cref{eq:lindeplinmap}.
\end{proof}

By combining the above property with the leftover hash lemma, we can now prove \cref{lem:subsetrank}.

\begin{proof}[Proof of \cref{lem:subsetrank}]
    Fix arbitrary sets $A,B\subseteq\F_2^n$ of size $2^k$.
    By \cref{lem:ellInftyLHL-new} and a union bound, there exists a linear map $L:\F_2^n\to\F_2^t$ with output length $t=k-C(1+\log n)$ that is surjective on both $A$ and $B$.
    In particular, there are subsets $A'\subseteq A$ and $B'\subseteq B$ of size $\binom{t}{\leq d/2}$ such that $L(A'),L(B')=\cB^t_{d/2}(0)$.
    Since $L$ is linear, we have that $L(A'+B')=L(A')+L(B')=\cB^t_d(0)$, and so
    \begin{equation*}
        \rank(\eval_d(A'+B'))\geq \rank(\eval_d(L(A'+B')))=\rank(\eval_d(\cB^t_d(0)))=\binom{t}{\leq d},
    \end{equation*}
    where we have invoked \cref{lem:ranklinmap}.
\end{proof}

Next, since it will be useful in \cref{sec:evasive}, we state here an alternative version of \cref{lem:subsetrank}, which is more helpful when $k$ is very small. Its proof follows in exactly the same manner as that of \cref{lem:subsetrank}, with the exception that we invoke \cref{lem:lhlsurj} instead of \cref{lem:ellInftyLHL-new}.
\begin{lemma}\label{lem:subsetrankLHL}
For all \(n\geq k\geq d\in\N\) (with \(d\) even) satisfying \(k>100\), the following holds. Let $A,B\subseteq\F_2^n$ be sets of size $2^k$.
    Then, there exist subsets $A'\subseteq A$ and $B'\subseteq B$ such that $|A'|,|B'|= \binom{k/100}{\leq d/2}$ and $\rank(\eval_d(A'+B'))\geq \binom{k/100}{\leq d}$.    
\end{lemma}

We now use \cref{lem:subsetrank} to prove \cref{thm:lowdegsumsetdisp}.

\begin{proof}[Proof of \cref{thm:lowdegsumsetdisp}]
    
    Without loss of generality, we may assume that $\X$ and $\Y$ are uniformly distributed over sets $A,B\subseteq\F_2^n$, respectively, each of size $2^k$.
    By \cref{lem:subsetrank}, there exist subsets $A'\subseteq A$ and $B'\subseteq B$ of size $\binom{t}{\leq d/2}$ such that
    \begin{equation*}
        \rank(\eval_d(A'+B'))\geq \binom{t}{\leq d},
    \end{equation*}
    with $t=k-C_0(1+\log n)$ for an absolute constant $C_0>0$.
    Since $A'+B'\subseteq A+B$, it follows that $f$ is constant on $A+B$ with probability at most
    \begin{equation*}
        2\cdot 2^{-\rank(\eval_d(A'+B'))}\leq 2^{-\binom{t}{\leq d}+1}.
    \end{equation*}
    By taking a union bound over all the at most
    \begin{equation*}
        \binom{2^n}{\binom{t}{\leq d/2}}^2\leq 2^{2n\binom{t}{\leq d/2}}
    \end{equation*}
    choices of $A'$ and $B'$, we conclude that the probability that $f$ is constant on some set $A+B$ is at most
    \begin{align}
        2^{2n\binom{t}{\leq d/2}}\cdot 2^{-\binom{t}{\leq d}+1}
        \leq 2^{2n\binom{k}{\leq d/2}+1}\cdot 2^{-\binom{t}{\leq d}}.\label{eq:boundFailProb}
    \end{align}

    We will now invoke \cref{lem:binom-large-t} with $\beta=\delta=1/2$, $k$ in place of $n$, and $t=k-C_0(1+\log n)$.
    Let $\gamma>0$ be the absolute constant correspoding to $\beta=\delta=1/2$ in \cref{lem:binom-large-t}.
    By hypothesis, we have $k\geq Cd n^{2/d}$ and $d\leq \frac{c\log n}{\log\log n}<k/2=\delta k$ for $C>0$ an appropriately large constant and $c>0$ an appropriately small constant.
    In particular, this means that we can enforce $0<c<1$ and choose $C>0$ large enough depending on $\gamma$ and $C_0$ so that $k\geq C d n^{\frac{2\log\log n}{\log n}}\geq C d \log n$ satisfies $t=k-C_0(1+\log n)\geq d$ and $t\geq (1-\gamma/d)k$.
    By \cref{lem:binom-large-t}, we then get that $\binom{t}{\leq d}\geq \frac{1}{2}\binom{k}{\leq d}$. Plugging this into \cref{eq:boundFailProb}, we conclude that the failure probability is at most
    \begin{align}
        2^{2n\binom{k}{\leq d/2}+1}\cdot 2^{-\frac{1}{2}\binom{k}{\leq d}}.\label{eq:boundFailProb2}
    \end{align}
    We also have
    \begin{equation*}
        2n\binom{k}{\leq d/2}\leq 2dn\left(\frac{2ek}{d}\right)^{d/2} \leq \frac{1}{4}(k/d)^d \leq \frac{1}{4}\binom{k}{\leq d},
    \end{equation*}
    where the second inequality uses that $k\geq Cd n^{2/d}$ for a sufficiently large constant $C>0$. Plugging this into \cref{eq:boundFailProb2}, we conclude that the probability that $f$ is constant on $A+B$ is at most
    \begin{equation*}
        2^{2n\binom{k}{\leq d/2}+1}\cdot 2^{-\frac{1}{2}\binom{k}{\leq d}} \leq 2^{-\frac{1}{4}\binom{k}{\leq d}+1}.\qedhere
    \end{equation*}
\end{proof}

\dobib

\subsection{Low-degree polynomials extract from sumset sources}\label{sec:lowdegsumsetext}

In this section, we prove \cref{thm:lowdegsumsetext} in full generality. We begin with the following definition of a special type of sumset sources, for which it will be easy to show that random low-degree polynomials extract.

\begin{definition}
    We say that a sumset source \(\W=\X+\Y\) has \emph{full $\eval_d$-rank} if the set $\eval_d(\supp(\X)+\supp(\Y))$
    is a collection of \(|\supp(\X)|\cdot|\supp(\Y)|\) linearly independent vectors.
\end{definition}

 We proceed with a simple proof that random low-degree polynomials extract from sumset sources with full eval-rank. Then, for the main part of the proof, we show that every sumset source is close to a convex combination of sumset sources with full eval-rank.

\begin{lemma}\label{lem:ext-full-rank}
    For any \(t,n\in\N\) and \(\eps>0\) such that \(t\geq64n/\eps^2\), the following holds with probability at least \(1-2\cdot2^{-\eps^2t^2/32}\) over the selection of a random degree \(d\) polynomial \(f:\F_2^n\to\F_2\). For any sets \(X,Y\subseteq\F_2^n\) of size \(t\) satisfying \(\rank(\eval_d(X+Y))=t^2\), we have that
    \[
    |\bias_{\X+\Y}(f)|\leq\eps,
    \]
    where \(\X\) and \(\Y\) are uniformly distributed over \(X\) and \(Y\), respectively.
\end{lemma}
\begin{proof}
    Fix any sets \(X,Y\subseteq\F_2^n\) of size \(t\) satisfying \(\rank(\eval_d(X+Y))=t^2\). By this rank condition, we know that the \(t^2\) random variables \(\{f(x+y)\}_{(x,y)\in X\times Y}\) are independent and uniform over \(\F_2\). Thus, if we define for every \((x,y)\in X\times Y\) the indicator random variable \(\mathbf{Z}_{x,y}:=1[f(x+y)=0]\), then
    \[
    \bias_{\X+\Y}(f)=2\left(\frac{1}{t^2}\sum_{x,y}\mathbf{Z}_{x,y}-\frac{1}{2}\right)=2\left(\frac{1}{t^2}\sum_{x,y}\mathbf{Z}_{x,y}-\E\left[\frac{1}{t^2}\sum_{x,y}\mathbf{Z}_{x,y}\right]\right),
    \]
    and a direct application of the Chernoff bound (\cref{lem:chernoff}) yields
    \[
\Pr_f\left[|\bias_{\X+\Y}|>\eps\right]\leq2e^{-\eps^2t^2/16}.
    \]
    Finally, by taking a union bound over all the at most \(\binom{2^n}{t}^2\leq2^{2nt}\) choices of \(X\) and \(Y\), we conclude that the probability that \(|\bias_{\X+\Y}(f)|>\eps\) for some sets \(X,Y\) is at most
    \[
    2^{2nt}\cdot2e^{-\eps^2t^2/16},
    \]
    which is at most \(2\cdot2^{-\eps^2t^2/32}\) when \(t\geq64n/\eps^2\), as desired.
\end{proof}

\noindent Next, we show every sumset source is close to a convex combination of sumset sources with full $\eval_d$-rank.

\begin{lemma}\label{lem:convexCombFullRank}
    There exists a constant $c>0$ such that for all \(n\geq k\geq d\in\N\), the following holds. Let \(\W=\X+\Y\) be an \((n,k)\)-sumset source. Then \(\W\) is $2^{-ck}$-close to a convex combination of flat sumset sources $\W^\star=\X^\star+\Y^\star$ with full eval\(_d\)-rank and such that $|\supp(\X^\star)|=|\supp(\Y^\star)|=\binom{k/6}{\lfloor d/2\rfloor}$.
\end{lemma}
\begin{proof}
    Let \(\Ext:\F_2^n\to\F_2^m\) be a linear map such that \(m=k/2\) and
    \begin{align}
        &\Ext(\X)\approx_{2\eps}\U_m\\
        &\Ext(\Y)\approx_{2\eps}\U_m. \label{eq:guarantees-LHL}
    \end{align}
    This map is guaranteed to exist by the leftover hash lemma (\cref{lem:lhl}) with $\eps=2^{-k/4}$ and a union bound. 
    Using the dependency reversal lemma (\cref{lem:dependency-reversal}), there exist functions $\Ext_X^{-1}$ and $\Ext_Y^{-1}$ and random variables $\A=(\A_0,\dots,\A_t)$ and $\B=(\B_0,\dots,\B_t)$ such that
\begin{align*}
    \X&\equiv\mathsf{Pick}\left(\Ext^{-1}_X(\Ext(\X),\A_0),\dots,\Ext^{-1}_X(\Ext(\X),\A_t);\U\right)\\
\Y&\equiv\mathsf{Pick}\left(\Ext^{-1}_Y(\Ext(\Y),\B_0),\dots,\Ext^{-1}_Y(\Ext(\Y),\B_t);\U^\pr\right),
\end{align*}
where \(\A_0,\dots,\A_t,\B_0,\dots,\B_t,\U,\U^\pr,\X,\Y\) are mutually independent, \(\U,\U^\pr\sim\{0,\dots,t\}\) and \(\mathsf{Pick}\) uses its last argument to pick among its first $t+1$ arguments (i.e., \(\mathsf{Pick}(v_0,v_1,\dots,v_t; i)\) outputs \(v_i\)). To see why the above is true, we consider an arbitrary fixing of \(\U,\U^\pr\) and simply apply the dependency reversal lemma. 

Now, by \cref{eq:guarantees-LHL} and the independence of $\X$ and $\Y$, there are random variables \(\mathbf{R},\mathbf{R}^\pr\sim\F_2^m\) independent of each other and the rest that are both uniformly random over \(\F_2^m\) and such that we can replace \(\Ext(\X),\Ext(\Y)\) with them. Recalling that \(\X,\Y\) are independent, we get from an application of the data-processing inequality and \cref{eq:guarantees-LHL} that
\begin{align}
\mathsf{Pick}\left(\Ext^{-1}_X(\Ext(\X),\A_0),\dots,\Ext^{-1}_X(\Ext(\X),\A_t);\U\right)&\approx_{2\eps}\mathsf{Pick}\left(\Ext^{-1}_X(\mathbf{R},\A_0),\dots,\Ext^{-1}_X(\mathbf{R},\A_t);\U\right),\nonumber\\
\mathsf{Pick}\left(\Ext^{-1}_Y(\Ext(\Y),\B_0),\dots,\Ext^{-1}_Y(\Ext(\Y),\B_t);\U^\pr\right)&\approx_{2\eps}\mathsf{Pick}\left(\Ext^{-1}_Y(\mathbf{R}^\pr,\B_0),\dots,\Ext^{-1}_Y(\mathbf{R}^\pr,\B_t);\U^\pr\right).\label{eq:replaceExt}
\end{align}

We now couple the randomness of \(\mathbf{R},\mathbf{R}^\pr\) in a specific way. 
Let \(\mathbf{L}\sim\F_2^{m\times m}\) be a uniformly random invertible matrix ($\mathbf{L}$ is obtained by sampling its $i$-th column uniformly at random from $\F_2^m$ conditioned on it being linearly independent of the previous $i-1$ columns).
Intuitively, we take appropriate disjoint subsets $\cB_0$ and $\cB_1$ of the radius-$d/2$ Hamming ball, and replace $\mathbf{R}$ and $\mathbf{R'}$ by applications of $\mathbf{L}$ to vectors in these sets.
More precisely, consider the sets $\cB_0,\cB_1\subseteq\F_2^m$ defined as
\begin{align*}
    &\cB_0 = \{u\in\F_2^m : \mathsf{wt}(u)=\lfloor d/2\rfloor, \supp(u)\subseteq\{1,\dots,m/3\}\},\\
    &\cB_1 = \{v\in\F_2^m : \mathsf{wt}(v)=\lfloor d/2\rfloor, \supp(v)\subseteq\{2m/3+1,\dots,m\}\}.
\end{align*}
Note that vectors in $\cB_0$ and $\cB_1$ are nonzero and have disjoint supports.
Moreover, $\cB_0+\cB_1$ is a subset of the radius-$d$ Hamming ball, and so $\rank(\eval_d(\cB_0+\cB_1))=|\cB_0|\cdot |\cB_1|$.

Let \(u_0,\dots,u_t\) be the elements in \(\mathcal{B}_0\), and let \(v_0,\dots,v_t\) be the elements in \(\mathcal{B}_1\). We argue that
\begin{align}
&\bigg(\mathsf{Pick}\left(\Ext^{-1}_X(\mathbf{R},\A_0),\dots,\Ext^{-1}_X(\mathbf{R},\A_t);\U\right),\mathsf{Pick}\left(\Ext^{-1}_Y(\mathbf{R}^\pr,\B_0),\dots,\Ext^{-1}_Y(\mathbf{R}^\pr,\B_t);\U^\pr\right)\bigg)\nonumber\\
&\approx_{m 2^{-m/3}} \bigg(\mathsf{Pick}\left(\Ext^{-1}_X(\mathbf{L}u_0,\A_0),\dots,\Ext^{-1}_X(\mathbf{L}u_t,\A_t);\U\right),\mathsf{Pick}\left(\Ext^{-1}_Y(\mathbf{L}v_0,\B_0),\dots,\Ext^{-1}_Y(\mathbf{L}v_t,\B_t);\U^\pr\right)\bigg)\nonumber\\
&:=(\X^\ast,\Y^\ast). \label{eq:couplerand}
\end{align}
To see why this holds, consider an arbitrary fixing of $(\U,\U')=(i,j)$.
Then, by an application of the data-processing inequality, it suffices to show that $(\mathbf{L}u_i,\mathbf{L}v_j)\approx_{m2^{-m/3}}(\mathbf{R},\mathbf{R'})$. Towards this end, recall that for any $i,j\in[t]$, the vectors $u_i$ and $v_j$ are nonzero with disjoint supports of size $m/3$ each. Let $\mathbf{L'}$ denote the $m\times (2m/3)$ matrix obtained by selecting columns of $\mathbf{L}$ indexed by the supports of $u_i$ and $v_j$.
Then, we have that $\mathbf{L'}\approx_{m 2^{-m/3}} \mathbf{M'}$, where $\mathbf{M'}$ is a uniformly random $m\times (2m/3)$ matrix.
To see this, note that a uniformly random vector in $\F_2^m$ will be linearly independent from any given collection of $2m/3$ vectors with probability at least $1-2^{-m/3}$, and then apply a union bound over all the $2m/3<m$ columns of $\mathbf{L'}$.
Therefore, letting $u'_i$ and $v'_j$ denote the restrictions of $u_i$ and $v_j$ to the coordinates in $\supp(u_i)\cup\supp(v_j)$, we have that $(\mathbf{L}u_i,\mathbf{L}v_j)\approx_{m 2^{-m/3}} (\mathbf{M'}u'_i,\mathbf{M'}v'_j)\equiv (\mathbf{R},\mathbf{R'})$.
The last step holds because $u'_i$ and $v'_j$ are linearly independent, so the random variables $\mathbf{M'}u'_i$ and $\mathbf{M'}v'_j$ are independent and uniformly distributed over $\F_2^m$.

We now analyze the $\eval_d$-rank of $\X^*+\Y^*$. Consider any fixing of the random variables \(\mathbf{L}\) and \(\A_0,\dots,\A_t,\B_0,\dots,\B_t\).
Upon such a fixing, \((\X^\ast,\Y^\ast)\) becomes of the form \((\X^\star,\Y^\star)\), where \(\X^\star,\Y^\star\) are independent and uniform over the sets
\begin{align*}
X^\star&:=\{\Ext^{-1}_X(Lu_0,a_0),\dots,\Ext^{-1}_X(Lu_t,a_t)\},\\
Y^\star&:=\{\Ext^{-1}_Y(Lv_0,b_0),\dots,\Ext^{-1}_Y(Lv_t,b_t)\},
\end{align*}
respectively. Then, notice that the support of \(\X^\star+\Y^\star\) is exactly
\[
    S^\star:=\{\Ext_X^{-1}(Lu_i,a_i)+\Ext_Y^{-1}(Lv_j,b_j)\}_{i,j\in[t]}.
\]
To analyze the $\eval_d$-rank of $S^\star$, recall from \cref{lem:ranklinmap} that applying linear transformations can only decrease the $\eval_d$-rank. Furthermore, note that for any \(i,j\in [t]\) it holds that
\[
    L^{-1}(\Ext(\Ext^{-1}_X(Lu_i,a_i) + \Ext^{-1}_Y(Lv_j,b_j)))=u_i+v_j.
\]
Furthermore, the composition \(L^{-1}\circ\Ext\) is linear, since \(L\) was linear (and invertible) and \(\Ext\) is also linear (since it comes from the leftover hash lemma). Thus,
\[
    \rank(\eval_d(S^\star))\geq\rank(\eval_d(L^{-1}(\Ext(S^\star))))=\rank(\eval_d(\{u_i+v_j\}_{i,j}))=\rank(\eval_d(\mathcal{B}_0+\mathcal{B}_1)).
\]
Since $|S^\star|\leq |\cB_0|\cdot|\cB_1|$, we get that $\rank(\eval_d(S^\star))=|\cB_0|\cdot|\cB_1|$, and so $\X^\star+\Y^\star$ has full $\eval_d$-rank.
Recalling \cref{eq:replaceExt,eq:couplerand} and the parameter settings $m=k/2$ and $\eps=2^{-k/4}$, this means that the sumset source $\X+\Y$ is $\eps^\star$-close to a convex combination of flat sumset sources $\X^\star+\Y^\star$ with full $\eval_d$-rank and support sizes $|\supp(\X^\star)|,|\supp(\Y^\star)|=\binom{m/3}{\lfloor d/2\rfloor} =\binom{k/6}{\lfloor d/2\rfloor}$, where $\eps^\star=4\eps+m2^{-m/3}\leq4\cdot2^{-k/4}+(k/2)\cdot2^{-k/6}\leq5k\cdot2^{-k/6}$. This is at most \(2^{-k/7}\) as long as \(k\) exceeds a big enough constant \(C\). Since the lemma is straightforward to obtain whenever \(k\leq C\),\footnote{This is because \(1-2^{-2k}\leq2^{-ck}\) whenever \(k\leq C\), provided that \(c>0\) is a small enough constant that depends on \(C\). Thus, in this regime, the lemma amounts to simply finding a sumset source \(\W^\star\) with full eval\(_d\)-rank that shares at least one support element with \(\W\), which is straightforward.} this completes the proof.
\end{proof}

\noindent Finally, we show how to combine \cref{lem:ext-full-rank} and \cref{lem:convexCombFullRank} to get \cref{thm:lowdegsumsetext}, restated here for convenience.

\lowdegsumsetext*
\begin{proof}
Fix any \(n\geq k\geq d\in\N\) and \(\eps>0\) such that \(k\geq Cd(n/\eps^2)^{1/\lfloor d/2\rfloor}\), where \(C\) is a sufficiently large constant,\footnote{What exactly ``sufficiently large'' means will become clear later in the proof.} and define \(\eps_0:=\eps/2\). By \cref{lem:convexCombFullRank}, every \((n,k)\)-sumset source \(\W=\X+\Y\) is \(2^{-c_0k}\)-close to a convex combination of sumset sources \(\W^\star=\X^\star+\Y^\star\) with full eval\(_d\)-rank and
\[
t:=|\supp(\X^\star)|=|\supp(\Y^\star)|=\binom{k/6}{\lfloor d/2\rfloor},
\]
where \(c_0>0\) is some constant. By the lower bound on \(k\) and the fact that \(C\) is a sufficiently large constant,
\[
t=\binom{k/6}{\lfloor d/2\rfloor}\geq\left(\frac{k/6}{\lfloor d/2\rfloor}\right)^{\lfloor d/2\rfloor}\geq(C/3)^{\lfloor d/2\rfloor}n/\eps^2=(C/3)^{\lfloor d/2\rfloor}n/(4\eps_0^2)\geq64n/\eps_0^2.
\]
Thus, \cref{lem:ext-full-rank} tells us that a random degree \(d\) polynomial \(f:\F_2^n\to\F_2\) is an \((\eps_0+2^{-c_0k})\)-extractor for \((n,k)\)-sumset sources with \(k\geq Cd(n/\eps^2)^{1/\lfloor d/2\rfloor}\), except with probability at most \(2\cdot2^{-\eps_0^2t^2/32}\).

We now argue that \(\eps_0+2^{-c_0k}\leq\eps\). For this, it suffices to argue that \(2^{-c_0k}\leq\eps_0\) (which was set to \(\eps/2\)). Thus, given the lower bound on \(k\), we just need \(2^{-c_0 Cd(n/\eps^2)^{1/\lfloor d/2\rfloor}}\leq\eps\). For this, it suffices to show \(2^{-c_0Cd(1/\eps)^{1/\lfloor d/2\rfloor}}\leq\eps\). And if \(C\) is a sufficiently large constant \(C\geq (\log_2 e)/(2c_0)\), then it suffices to show \(2^{-(\log_2e)(d/2)(1/\eps)^{1/\lfloor d/2\rfloor}}\leq\eps\), or \(e^{\lfloor d/2\rfloor(1/\eps)^{1/\lfloor d/2\rfloor}}\geq1/\eps\), which is \(e^{(1/\eps)^{1/\lfloor d/2\rfloor}}\geq(1/\eps)^{1/\lfloor d/2\rfloor}\). And this follows from the standard inequality \(x\leq e^x\) for all real \(x\).

Finally, it is easy to verify that \(2\cdot2^{-\eps_0^2t^2/32}\leq2^{-\eps^2\binom{k/C}{2\lfloor d/2\rfloor}}\), using our definitions of \(\eps_0\) and \(t\) above, and the fact that \(k\geq Cd(n/\eps^2)^{1/\lfloor d/2\rfloor}\) for a sufficiently large constant \(C\).
\end{proof}

\dobib

\subsection{Impossibility results}\label{sec:impossibility}
Above, we proved that random low-degree polynomials are good extractors for sumset sources (\cref{thm:lowdegsumsetext}). Given this result, it is natural to ask just how tight it is: in particular, can one show that every low-degree polynomial is nevertheless constant on some relatively large sumset source? As it turns out, the work of Cohen and Tal \cite{CT15} already gives an answer to this question. Indeed, they show that every low-degree polynomial is constant on some relatively large \emph{affine source}, which is a special case of a sumset source. Thus, their result already implies that every degree \(d\) polynomial \(f:\F_2^n\to\F_2\) is constant on some \((n,k)\)-sumset source \(\W\) with \(k\geq\Omega(dn^{\frac{1}{d-1}})\), proving that \cref{thm:lowdegsumsetext} is tight up to the constant $2$ in the exponent.

In this section, we instead ask whether such a result can be obtained for \emph{two independent sources} -- which can be viewed as the \emph{other} well-known specialization of sumset sources. We prove such a result below, and show that it holds even if we consider the following (more general) family of functions \(f:\F_2^n\times\F_2^n\to\F_2\).

\begin{definition}[Left-degree]\label{def:leftdeg}
We say that a function $f:\F_2^n\times\F_2^n\to\F_2$ has \emph{left-degree $d$} if for any fixed $y\in\F_2^n$, the restricted function $f(\cdot,y)$ can be written as an $\F_2$-polynomial of degree $\leq d$.
\end{definition}

Note that every polynomial of degree \(\leq d\) also has left degree \(\leq d\), but a function with left degree \(\leq d\) can be much more complex than a degree \(\leq d\) polynomial. Now, given this definition, we prove the following impossibility result for two-source dispersers that are computed by functions with low left-degree.

\begin{theorem}\label{thm:dispLB}
    There is a constant $c>0$ such that for every positive integers $n$ and $d\leq c\log n$, the following holds.
    For any function $f:\F_2^n\times\F_2^n\to\F_2$ with left-degree $\leq d$, there exist independent $(n,k)$-sources $\X,\Y$ with $k\geq c d n^{1/d}$ such that $f(\X,\Y)$ is constant.
\end{theorem}

Before we prove \cref{thm:dispLB}, we present a corollary for degree $\leq d$ polynomials.

\begin{corollary}
There is a constant $c>0$ such that for every positive integers $n$ and $d\leq c\log n$, the following holds. For every polynomial $f:\F_2^{2n}\to\F_2$ with degree $\leq d$, there exist independent $(n,k)$-sources $\X,\Y$ with $k\geq c d n^{\frac{1}{d-1}}$ such that $f(\X,\Y)$ is constant.
\end{corollary}
\begin{proof}
Every polynomial $f:\F_2^{2n}\to\F_2$ of degree at most $d$ can be written as
\begin{equation*}
    f(x,y) = g(x,y) + h(x),
\end{equation*}
such that $g$ has degree at most $d$ and does not feature monomials of $f$ containing only variables from $\{x_1,\dots,x_n\}$, and $h$ collects all remaining monomials.
Note that $g$ has left-degree at most $d-1$.
Thus, by \cref{thm:dispLB}, there exist independent $(n,k)$-sources $\X,\Y$ with $k\geq c d n^{\frac{1}{d-1}}$ such that $g(\X,\Y)$ is constant.

Next, without loss of generality, assume that $\Pr[h(\X)=0]\geq 1/2$.
In this case, the random variable $\X'=(\X\mid h(\X)=0)$ satisfies
\begin{equation*}
    \minH(\X')\geq \minH(\X)-1\geq k-1\geq c' d n^{\frac{1}{d-1}}
\end{equation*}
for some absolute constant $c'>0$.
Since $f(\X',\Y)=g(\X',\Y)$ is constant, the desired result follows.
\end{proof}

Now, in order to prove \cref{thm:dispLB}, we revisit the argument used by Cohen and Tal~\cite{CT15} in the context of affine sources. In particular, we will use the following result from~\cite[Appendix B]{CT15}, which is a consequence of the Chevalley-Warning theorem. It says that if a low-degree polynomial $f$ vanishes on a subspace $V$, then it also vanishes on many shifts of that subspace.

\begin{lemma}[\protect{\cite[Appendix B]{CT15}}]\label{lem:cohental}
    Consider a subspace $V\subseteq\F_2^n$ of dimension $t$ and a polynomial $f:\F_2^n\to\F_2$ of degree at most $d$ such that $f(v)=0$ for all $v\in V$.
    Then, it holds that
    \begin{equation*}
        \left|\{x\in\F_2^n:\forall v\in V, f(x+v)= 0\}\right| \geq 2^{n-\sum_{j=0}^{d-1}(d-j)\binom{t}{j}}.
    \end{equation*}
\end{lemma}

With this tool in hand, we are ready to prove \cref{thm:dispLB}.

\begin{proof}[Proof of \cref{thm:dispLB}]
    We may alternatively see a function $f:\F_2^n\times\F_2^n\to\F_2$ of left-degree at most $\leq d$ as a collection of $2^n$ polynomials $\{f_y:\F_2^n\to\F_2\}_{y\in\F_2^n}$, each of degree $\leq d$, where $f_y(x):=f(x,y)$.
    Without loss of generality, we assume that $f_y(0)=0$ for at least $2^{n-1}$ choices of $y$.
    The desired result follows if we can find sets $X,Y\subseteq\F_2^n$, each of size at least $2^{c d n^{1/d}}$, such that $f_y(x)=0$ for every $y\in Y$ and $x\in X$. Indeed, we can then take $\X$ and $\Y$ to be uniformly distributed over $X$ and $Y$, respectively.
    
    Towards this end, consider sampling a subspace $V\subseteq\F_2^n$ of dimension
    \begin{equation*}
        t=\left\lfloor\frac{d}{e}\left(\frac{en}{4d^3}-1\right)^{1/d}\right\rfloor
    \end{equation*}
    by iteratively sampling $v_i$ uniformly at random from $\F_2^n\setminus\spanvec(v_1,\dots,v_{i-1})$ for $i=1,\dots,t$.
    Looking ahead, we will choose $c>0$ small enough (i.e., smaller than $\frac{1}{4\log(2 e)}$) so that $d\leq \frac{t-1}{2}$ for $d\leq c \log n$.
    
    Using \cref{lem:cohental}, the probability that some fixed polynomial $f:\F_2^n\to\F_2$ of degree at most $d$ with $f(0)=0$ satisfies $f(x)=0$ for all $x\in V$ is at least
    \begin{align}
        \prod_{i=0}^{t-1} \frac{2^{n-\sum_{j=0}^{d-1}(d-j)\binom{i}{j}}-2^i}{2^n} &\geq \prod_{i=0}^{t-1} \frac{2^{n-1-\sum_{j=0}^{d-1}(d-j)\binom{i}{j}}}{2^n}\label{eq:prodineq1}\\
        &= 2^{-t-\sum_{i=0}^{t-1}\sum_{j=0}^{d-1}(d-j)\binom{i}{j}}\nonumber\\
        &\geq 2^{-t-d^2 t (et/d)^{d-1}}\label{eq:prodineq2}\\
        &\geq 2^{-n/2}.\label{eq:prodineqfinal}
    \end{align}
    \cref{eq:prodineq1} holds because
    \begin{equation*}
        n-\sum_{j=0}^{d-1}(d-j)\binom{t-1}{j} \geq n - d^2 \left(\frac{et}{d}\right)^{d-1} \geq 3n/4\geq 2t,
    \end{equation*}
    which follows from the fact that $d\leq \frac{t-1}{2}$ and by our choice of $t$.
    \cref{eq:prodineq2} uses the fact that
    \begin{equation*}
        \sum_{i=0}^{t-1}\sum_{j=0}^{d-1}(d-j)\binom{i}{j}\leq d^2 t \left(\frac{et}{d}\right)^{d-1}< n/4,
    \end{equation*}
    again by our choice of $t$.

    Now, if $\A\sim[2^n]$ denotes the number of polynomials in the collection $\{f_y:y\in\F_2^n, f_y(0)=0\}$ such that $f_y(v)= 0$ for all $v\in V$, then by linearity of expectation and \cref{eq:prodineqfinal}, we have
    \begin{equation*}
        \E[\A]\geq 2^{n-1}\cdot 2^{-n/2}=2^{n/2-1}\geq 2^t.
    \end{equation*}
    Therefore, by an averaging argument, there exist a $t$-dimensional subspace $V$ and a set $Y\subseteq \F_2^n$ of size at least $2^t$ such that $f_y(V)\equiv 0$ for all $y\in Y$. Taking $X=V$ and this $Y$ (and $\X$ and $\Y$ uniformly distributed over these subsets, respectively) concludes the argument.
\end{proof}

\dobib

\section{Low-degree polynomials yield evasive sets}\label{sec:evasive}

\subsection{Low-degree, low-error two-source extractors}

Recall that in \cref{thm:lowdegsumsetext}, we showed that a random degree \(d\) polynomial \(f:\F_2^n\to\F_2\) is an \(\eps\)-extractor for \(n\)-bit sumset sources with min-entropy \(k=O(d(n/\eps^2)^{1/\lfloor d/2\rfloor})\). While this can handle any range of parameters, notice that in order to achieve ``low error'' \(\eps=k^{-\omega(1)}\), it requires superconstant degree \(d=\omega(1)\). It is natural to ask whether low error can still be achieved with polynomials of constant degree \(d=O(1)\).

As a step in this direction, in this section, we show the existence of a low-error two-source extractor for arbitrary linear min-entropy $k=\delta n$, which is computable by a degree $\leq 4$ polynomial. In contrast, the best known explicit constructions of low-error two-source extractors require min-entropy $k\approx 0.44n$~\cite{Bou05,Lew19}. More precisely, we prove the following.

\begin{restatable}[Low-degree low-error two-source extractor]{theorem}{lowErrorTwoExt}\label{thm:lowErrorLowDeg2Ext}
    For every $\delta>0$ there is some $\zeta>0$ such that there exists a $(k=\delta n, \eps=2^{-\zeta n})$-two-source extractor $\Ext:\F_2^n\times\F_2^n\to\F_2$ that is computable by an $\F_2$-polynomial of degree $\leq 4$.
    Moreover, such a function $\Ext$ can be sampled with probability $0.99$ in time $O(n^3)$, and $\Ext(x,y)$ can be evaluated in time $O(n^3)$ for every input $(x,y)$.
\end{restatable}

We prove \cref{thm:lowErrorLowDeg2Ext} by combining a previously known connection between subspace-evasive sets (over $\F_2$) and low-error two-source extractors for linear min-entropy, and a construction of such subspace-evasive sets from degree-$2$ polynomials.
This connection was originally conditional on (a weak form of) the \emph{approximate duality conjecture}~\cite{BR15}.
However, this conjecture is now a theorem since it is implied by the Polynomial Freiman-Ruzsa theorem, which was recently proved by Gowers, Green, Manners, and Tao~\cite{GGMT23}.

\subsubsection*{A known reduction to low-degree subspace-evasive sets}

We make the connection clear here for completeness. 
We need the following definition and previously known results.

\begin{definition}[Rank of a function]
    We say that a function $f:\F_2^n\times\F_2^n\to\F_2$ has \emph{rank} at most $r$ if there exist functions $g_1,g_2:\F_2^n\to\F_2^r$ such that
    \begin{equation*}
        f(x,y) = \langle g_1(x),g_2(y)\rangle.
    \end{equation*}
    Equivalently, $f$ has rank at most $r$ if and only if its $2^n\times 2^n$ truth table has rank at most $r$ over $\F_2$.
\end{definition}
A uniformly random function will have exponentially large rank with high probability.
The following result states that every \emph{linear-rank} two-source disperser for linear min-entropy is also a low-error two-source extractor for essentially the same min-entropy.

\begin{theorem}[\protect{\cite[Lemma 2.17]{BR15} and Polynomial Freiman-Ruzsa~\cite{GGMT23}}]\label{thm:DispToExt}
    For every $C,\gamma,\rho>0$ there exists $\eta>0$ such that the following holds for all $n$: Let $f:\F_2^n\times\F_2^n\to\F_2$ be a $(k=\rho n)$-two-source disperser with rank at most $Cn$.
    Then, $f$ is also a $(k'=(\rho+\gamma)n,\eps = 2^{-\eta n})$-two-source extractor.
\end{theorem}

Ben-Sasson, Lovett, and Ron-Zewi~\cite{BLR12} obtained an entropy-error tradeoff for this result, which allows one to drop the min-entropy $k'$ to $O(n/\log n)$ while having error $2^{-\Omega(\sqrt{n})}$.
For the sake of exposition, we do not use this alternative version, but remark that it can be applied here as well.

Based on the theorem above, our goal is to construct a linear-rank two-source disperser for any sublinear min-entropy.
We use a simple observation already present in~\cite{PR04,BR15}, whose short proof we include for completeness.

\begin{lemma}[\cite{PR04,BR15}]\label{lem:dim}
    If $A,B\subseteq\F_2^n$ are sets such that\footnote{We take $\dim A$ to mean $\dim \spanvec(A)$.} $\dim A + \dim B > n+1$, then
    \begin{equation*}
        \langle A,B\rangle := \{\langle a, b \rangle: a\in A, b\in B\} = \{0,1\}.
    \end{equation*}
\end{lemma}
\begin{proof}
    First, suppose that $\langle A,B\rangle =\{0\}$.
    This means that $B$ lies in the subspace orthogonal to $\spanvec(A)$, and so $\dim A + \dim B \leq n$.
    On the other hand, if $\langle A,B\rangle =\{1\}$, then fix any $a\in A$ and consider instead the set $A'=A-a$.
    Note that $\langle A', B\rangle =\{0\}$ and that $\dim A' \geq \dim A - 1$.
    By the argument above, we must have $\dim A' + \dim B \leq n$, and so also $\dim A + \dim B \leq n+1$.
\end{proof}

Given an input length $n$ and min-entropy requirement $k\leq n$, \cref{lem:dim} suggests building an encoding function $h:\F_2^n\to\F_2^r$ that is subspace-evasive in the sense of Pudlák and R\"odl~\cite{PR04}.

\begin{definition}[Subspace-evasive sets and functions]
    A set $S\subseteq \F_2^r$ is said to be \emph{$(\ell,2^k)$-subspace-evasive} if $|S\cap V|< 2^k$ for any dimension-$\ell$ subspace $V$ of $\F_2^r$.
    An injective function $h:\F_2^n\to\F_2^r$ is \emph{$(\ell,2^k)$-subspace-evasive} if $h(\F_2^n)$ is $(\ell,2^k)$-subspace-evasive.
\end{definition}

Given a function $h:\F_2^n\to\F_2^r$, we consider the two-source extractor
\begin{equation*}
    \Ext_h(x,y) = \langle h(x),h(y)\rangle.
\end{equation*}
The following lemma relates the subspace-evasiveness of $h$ to the dispersion and rank of $\Ext_h$.
\begin{lemma}\label{lem:IPlinrank}
    Given an injective $(\ell = r/2, 2^k)$-subspace-evasive function $h:\F_2^n\to\F_2^r$, it follows that $\Ext_h$ is a $k$-two-source disperser of rank at most $r$.    
\end{lemma}
\begin{proof}
    Fix any two sets $A,B\subseteq\F_2^n$ of size $2^k$ each.
    We have that $\dim h(A),\dim h(B) \geq r/2+1$, since $h$ is $(r/2,2^k)$-subspace-evasive. In particular, this means that $\dim h(A) + \dim h(B) > r+1$.
    By \cref{lem:dim}, we conclude that $\Ext_h(A,B)=\langle h(A),h(B)\rangle=\zo$.
\end{proof}

Combining \cref{lem:IPlinrank} with \cref{thm:DispToExt}, we immediately obtain the following theorem, which reduces constructing a low-error two-source extractor for any linear min-entropy to constructing an appropriate subspace evasive set.
\begin{theorem}\label{thm:SubEvasiveToExt}
    For every $C,\gamma,\rho>0$ there exists $\eta>0$ such that the following holds for all $n$:
    Let $h:\F_2^n\to\F_2^r$ be an injective $(\ell = r/2,2^k=2^{\rho n})$-subspace-evasive function with $r\leq Cn$.
    Then, $\Ext_h(x,y)=\langle h(x),h(y)\rangle$ is a $(k'=(\rho+\gamma)n,\eps=2^{-\eta n})$-two-source extractor.
    Moreover, if $h$ is explicit, then so is $\Ext_h$.
\end{theorem}

\subsubsection*{A random construction of low-degree subspace-evasive sets}

With \cref{thm:SubEvasiveToExt} in mind,
it remains to construct an $(\ell,2^k)$-subspace-evasive function $h:\F_2^n\to\F_2^{n+r}$ with $r=O(n)$, $k=o(n)$, and $\ell=\frac{n+r}{2}$.
Furthermore, we would like to construct $h$ in a way that $\Ext_h$ is computable by a low-degree $\F_2$-polynomial.
Towards this end, we consider the function
\begin{equation*}
    h(x) = (x_1,\dots,x_n,f_1(x), f_2(x),\dots, f_r(x)),
\end{equation*}
where $f_1,\dots,f_r:\F_2^n\to\F_2$ are random degree $d$ polynomials.
\begin{theorem}\label{thm:dimexpand}
    Let $k = 10d n^{1/d}$ and $r=11n$.
    Then, with probability at least $1-2^{-3n^2}$ over the sampling of $f_1,\dots,f_r$ we have that $h$ is an injective $(\ell=\frac{n+r}{2},2^k)$-subspace-evasive function.
\end{theorem}

Sudakov and Tomon~\cite{ST23} also consider subspace-evasive sets induced by collections of multivariate polynomials, but they analyze instead the setting where the field is large and both the dimension and intersection size are small. Next, we need the following lemma.
\begin{lemma}[\protect{\cite[Lemma 1.4]{BHL12} and \cite[Theorem 1.5]{KS05}, adapted}]\label{lem:fullevalrank}
    For any set $S\subseteq\F_2^n$ of size $2^k$ we have that $\rank(\eval_d(S))\geq \binom{k}{\leq d}$.
\end{lemma}

With this lemma in hand, we are ready to prove \cref{thm:dimexpand}.

\begin{proof}[Proof of \cref{thm:dimexpand}]

The injectivity of $h$ holds by construction, so we focus on the subspace-evasiveness of $h$.
We need to show that with high probability it holds that for any set $A\subseteq \F_2^n$ of size $2^k$ we have $\dim h(A) > \frac{r+n}{2}$.
We may write $h(x) = (x,M\eval_d(x))$, where $M$ is an $r\times \binom{n}{\leq d}$ matrix with i.i.d.\ uniformly random $\F_2$ entries.

Fix such a set $A$. Then, by \cref{lem:fullevalrank}, there exists a subset $S\subseteq A$ of size $r=10n\leq \binom{k}{\leq d}$ such that the vectors in $\eval_d(S)$ are all linearly independent.
It suffices to show that $\dim (M\eval_d(S)) > \frac{r+n}{2}=5.5n$.
    Choose $\frac{n+r}{2}$ vectors in $\eval_d(S)=\{v_1,v_2,\dots,v_r\}$.
    Without loss of generality, we may take these vectors to be $v_1,\dots,v_{\frac{n+r}{2}}$.
    Since the $v_i$'s are linearly independent, the random variables $(Mv_i)_{i\in[r]}$ are also independent.
    For a fixed $i>\frac{n+r}{2}$, the probability that $Mv_i$ lies in $\spanvec(Mv_1,\dots,Mv_{\frac{n+r}{2}})$ is $2^{-\frac{r-n}{2}}$.
    Therefore, the probability that $Mv_{\frac{n+r}{2}+1},\dots,Mv_r$ all lie in this subspace is
    \begin{equation*}
        (2^{-\frac{r-n}{2}})^{r-\frac{n+r}{2}} = 2^{-25n^2}.
    \end{equation*}
    Then, a union bound over the $\binom{|S|}{\frac{n+r}{2}} = \binom{r}{\frac{n+r}{2}}$ choices for the initial set of $\frac{n+r}{2}$ vectors from $\eval_d(S)$ shows that the probability that $\dim h(S)\leq \frac{n+r}{2}$ is at most
    \begin{equation*}
        \binom{r}{\frac{n+r}{2}}\cdot 2^{-25n^2} \leq 2^r \cdot 2^{-25n^2}= 2^{11n-25n^2} \leq 2^{-14n^2}.
    \end{equation*}
    Finally, we take a union bound over all the $\binom{2^n}{r}$ choices of $S$ to conclude that the probability that $h$ is not subspace-evasive is at most
    \begin{equation*}
        \binom{2^n}{r} \cdot 2^{-14n^2} \leq 2^{nr - 14n^2} = 2^{-3n^2}.\qedhere
    \end{equation*}
\end{proof}

We can now combine \cref{thm:dimexpand} for degree $d=2$ with \cref{thm:SubEvasiveToExt} to immediately obtain a low-error two-source extractor for arbitrary linear min-entropy, which in particular yields \cref{thm:lowErrorLowDeg2Ext}.
\begin{corollary}
    Set $r=11n$.
    Sample $r$ random degree $2$ polynomials $f_1,\dots,f_r:\F_2^n\to\F_2$ and define $h:\F_2^n\to\F_2^{n+r}$ as $h(x) = (x_1,\dots,x_n,f_1(x),\dots,f_r(x))$.
    Then, with probability at least $1-2^{-3n^2}$ over the sampling of $f_1,\dots,f_r$ the following holds:
    For any $\delta >0$ there exists $c>0$ such that the function $\Ext_h:\F_2^n\times\F_2^n\to\F_2$ given by
    \begin{equation*}
        \Ext_h(x,y) = \langle h(x), h(y)\rangle
    \end{equation*}
    is a $(k' = \delta n,\eps=2^{-c n})$-two-source extractor.
    Furthermore, $\Ext_h$ is computable by a $2n$-variate $\F_2$-polynomial of degree at most $4$.
\end{corollary}

\subsection{Improved impossibility results for dispersing from polynomial and variety sources}

In this section, we show how to construct what we call \emph{sumset-evasive sets} using low-degree polynomials.
Then, we show how our constructions yield new impossibility results for dispersing from polynomial and variety sources using sumset dispersers, improving and simplifying results of Chattopadhyay, Goodman, and Gurumukhani~\cite{CGG24}.

\begin{definition}[Sumset-evasive sets and functions]
    A set \(S\subseteq\F_2^r\) is said to be \emph{\(\ell\)-sumset-evasive} if for any sets \(A,B\subseteq\F_2^r\) each of size \(2^\ell\) it holds that $A+B\not\subseteq S$. We say that a function \(f:\F_2^n\to\F_2^r\) is \(\ell\)-sumset-evasive if the image \(f(\F_2^n)\subseteq\F_2^r\) is \(\ell\)-sumset-evasive.
\end{definition}

We show that the same construction we used to obtain low-degree subspace-evasive sets above also works for sumset-evasiveness, if we use our new \cref{lem:subsetrankLHL} instead of \cref{lem:fullevalrank}. To keep things simple, we focus here on even degree \(d\), and note that it is straightforward to extend our results to hold for all degrees \(d\), at a slight loss in parameters (as was done for our fully general result on sumset extractors in \cref{sec:lowdegsumsetext}).

\begin{theorem}\label{thm:sumsetevasive}
There exist constants $C,c>0$ such that the following holds for all positive integers $k\geq t\geq 200d$ with \(d\) even and $\binom{t/100}{\leq d/2}\geq 4d^2 (2e)^d$.
Let \(f_1,\dots,f_r:\F_2^k\to\F_2\) be independent random degree \(d\) polynomials with
\begin{equation*}
    r\geq \frac{8d^2(2e)^d k}{\binom{t/100}{\leq d/2}}.
\end{equation*}
Consider the function \(h:\F_2^k\to\F_2^{k+r}\) defined as \(h(x)=(x,f_1(x),\dots,f_r(x))\). Then
\[
    \Pr_{f_1,\dots,f_r}\left[h(\F_2^k)\text{ is not $t$-sumset-evasive}\right]\leq 2^{-\frac{r}{2}\cdot\binom{t/100}{\leq d}}.
\]
\end{theorem}

\begin{proof}
Let \(S\subseteq\F_2^{k+r}\) denote the image of \(h\), and recall it is of the form
\[
S=\{(x,f(x)) : x\in\F_2^k\},
\]
where \(f:\F_2^k\to\F_2^r\) is defined as \(f(x):=(f_1(x),\dots,f_r(x))\). First, suppose there exist \(A,B\subseteq\F_2^n\) each of size at least \(2^t\) such that \(A+B\subseteq S\). Since the first \(k\) bits of \(S\) determine the rest, the same is true of both $A$ and $B$. In particular, there must exist some \(A^\pr,B^\pr\subseteq\F_2^k\) each of size at least \(2^t\) and functions \(\phi,\psi:\F_2^k\to\F_2^r\) such that 
\begin{align*}
A&=\{(a,\phi(a)):a\in A^\pr\},\\
B&=\{(b,\psi(b)):b\in B^\pr\}.
\end{align*}
Furthermore, again because \(A+B\subseteq S\), we know that for every \(a\in A^\pr,b\in B^\pr\),
\[
(a,\phi(a))+(b,\psi(b))=(a+b,\phi(a)+\psi(b))=(a+b,f(a+b)).
\]
In particular, this implies that \(f(a+b)=\phi(a)+\psi(b)\) for every \(a\in A^\pr,b\in B^\pr\).

Next, since \(A^\pr,B^\pr\) each have size at least \(2^t\), we know by \cref{lem:subsetrankLHL} that there must exist some \(A^\prpr\subseteq A^\pr,B^\prpr\subseteq B^\pr\) each of size \(\tau= \binom{t/100}{\leq d/2}\) such that 
\begin{equation}\label{eq:fullsumsetrank}
\rank(\eval_d(A^\prpr+B^\prpr))\geq \binom{t/100}{\leq d} \geq (2e)^{-d} d^{-2} \tau^2 =: \tau'.
\end{equation}
The rightmost inequality uses the fact that $\binom{t/100}{\leq d}\geq \left(\frac{t}{100d}\right)^d$ and that $\tau^2 =\binom{t/100}{\leq d/2}^2 \leq d^2\left(\frac{2et}{100d}\right)^d$, which in turn use the standard inequalities $\binom{u}{\leq v}\geq \binom{u}{v}\geq (u/v)^v$ and $\binom{u}{\leq v}\leq v \binom{u}{v}\leq v\left(eu/v\right)^v$, valid for $v\leq u/2$.

Furthermore, if we define \(\phi^\prpr:A^\prpr\to\F_2^r,\psi^\prpr:B^\prpr\to\F_2^r\) to denote the restrictions of \(\phi,\psi\) to \(A^\prpr,B^\prpr\) respectively, it still holds that \(f(a+b)=\phi^\prpr(a)+\psi^\prpr(b)\) for all \(a\in A^\prpr,b\in B^\prpr\).
Now, note that by \cref{eq:fullsumsetrank} there are at least $\tau'$ choices of pairs \(a\in A^\prpr,b\in B^\prpr\) such that the corresponding random variables \(f(a+b)\) are independent and uniformly distributed over \(\F_2^r\), while \(\phi^\prpr(a)+\psi^\prpr(b)\) is a fixed value. Thus, the probability that \(f(a+b)=\phi^\prpr(a)+\psi^\prpr(b)\) is precisely \(2^{-r}\).
Therefore, by the independence of $f(a+b)$ for the choices of $a\in A^\prpr$ and $b\in B^\prpr$ above, the probability they are equal for \emph{every} fixed pair \(a\in A^\prpr,b\in B^\prpr\) above is precisely \(2^{-r\cdot \tau'}\). 
Applying a union bound over all possible choices of \(A^\prpr,B^\prpr,\phi^\prpr,\psi^\prpr\) yields that the probability that such objects exist is at most
\begin{equation*}
    2^{-r\tau'}\cdot\binom{2^k}{\tau}^2\cdot((2^r)^\tau)^2\leq2^{-r\tau'+2k\tau+2r\tau},
\end{equation*}
which is at most \(2^{-r\tau'/2}\) provided that $r\geq \frac{4\tau k}{\tau'-4\tau}=\frac{4k}{(2e)^{-d}d^{-2}\tau-4}$, the result follows by noting that our hypotheses implies $(2e)^{-d}d^{-2}\tau-4\geq (2e)^{-d}d^{-2}\tau/2$, so it suffices to choose any $r\geq 8d^2(2e)^d k/\tau$.
\end{proof}

Note that for the functions $h$ from \cref{thm:sumsetevasive} we have that $\ZZ= h(\U_k)$ is a degree $\leq d$ polynomial source of length $n=k+r$ and min-entropy $k$ such that, with high probability over the choice of $h$, it holds that $\supp(\ZZ)$ does not contain any sumset $A+B$ with $|A|,|B|\geq 2^t$.
Moreover, the \emph{entropy gap} of $\ZZ$ is 
\begin{equation*}
    n-k=r=O\left(\frac{8d^2(2e)^d k}{\binom{t/100}{\leq d/2}}\right)=O\left(8d^2(2e)^d n\left(\frac{100d}{2t}\right)^{d/2}\right) \leq n\left(\frac{Cd}{t}\right)^{d/2},
\end{equation*}
for a sufficiently large constant $C>0$.
The third equality uses the fact that $k\leq n$ and $\binom{a}{\leq b}\geq \binom{a}{b}\geq \left(\frac{a}{b}\right)^{b}$.
Using this and setting $t=\alpha\log n$ for a sufficiently small constant $\alpha>0$ yields the following corollary of \cref{thm:sumsetevasive}, which states that there exist polynomial NOBF sources\footnote{Following~\cite[Definition 4]{CGG24}, a degree $d$ polynomial NOBF source $\ZZ\sim\F_2^n$ with min-entropy $k$ is a degree $d$ polynomial source with the following structure: There exists a set $G\subseteq[n]$ of $k$ good coordinates which are independent and uniformly distributed, and each coordinate outside $G$ is computed by a degree $\leq d$ polynomial of the coordinates in $G$. Polynomial NOBF sources are a special case of both polynomial sources and variety sources.} with extremely high min-entropy that avoid even very tiny sumsets.
\begin{corollary}\label{coro:improveCGG}
    For any constant $c>0$ there exists a constant $C>0$ such that the following holds for all positive integers $n\geq C$ and even degrees $d$. There exists a degree $d$ polynomial NOBF source $\ZZ\sim \F_2^n$ with min-entropy $\minH(\ZZ)\geq n - n\left(\frac{Cd}{\log n}\right)^{d/2}$ such that $\supp(\ZZ)$ does not contain any sumset $A+B$ with $|A|,|B|\geq n^c$.
\end{corollary}

\cref{coro:improveCGG} generalizes~\cite[Theorem 5.11]{CGG24}, which worked for $d=2$ only, to all even degrees $d$, and it has a simpler proof. It also yields the following corollary, extending the impossibility result from~\cite[Section 5.3]{CGG24} on using sumset dispersers to disperse from polynomial NOBF sources in a black-box manner to larger degrees $d>2$.

\begin{corollary}[Sumset dispersers cannot disperse from high-entropy polynomial NOBF sources and variety sources, any even degree $d$]\label{coro:impdisp}
    Fix any even degree $d\in\N$.
    Then, there exists a constant $C>0$ such that for any integer $n\geq C$ the following holds.
    Sumset dispersers cannot be used to disperse from degree $d$ polynomial NOBF sources over $\F_2^n$ with min-entropy $k=n-n\left(\frac{Cd}{\log n}\right)^{d/2}$ in a black-box manner.
    In particular, this also means that sumset dispersers cannot be used to disperse from degree $d$ variety sources with the same min-entropy $k$ in a black-box manner.
\end{corollary}
\begin{proof}
    In order to be able to conclude that $k'$-sumset disperser is also a disperser for another class of sources $\cC$ using only its black-box property (i.e., that it is not constant on any sumset $A+B$ with $|A|,|B|\geq 2^{k'}$), one needs that for every $\X\in\cC$, $\supp(\X)$ contains a sumset $A+B$ with $|A|,|B|\geq 2^{k'}$.
    Since we can take $k'=\alpha\log n$ for an arbitrarily small constant $\alpha>0$ and there are no sumset dispersers for such $k'$ (because every function is constant on an $\Omega(\log n)$-dimensional subspace, and affine sources are sumset sources), the result follows.
    To see the ``In particular'' part, it suffices to note that degree $d$ polynomial NOBF sources with min-entropy $k$ are also degree $d$ variety sources with min-entropy $k$~\cite[Claim 1]{CGG24}.
\end{proof}

\dobib

\section{Open problems}\label{sec:conclusions}

We list here some of our favorite directions for future research:
\begin{itemize}
    \item In \cref{thm:lowdegsumsetext}, we showed that most degree $\leq d$ polynomials are sumset dispersers (in fact, extractors) for min-entropy $k=O(dn^{1/\lfloor d/2\rfloor})$. On the other hand, we also know that no degree $\leq d$ polynomial is a sumset disperser for min-entropy $k=c\cdot dn^{1/(d-1)}$, where \(c>0\) is some constant. Can we narrow this gap?

    \item We conjecture that most degree $\leq d$ polynomials are sumset extractors with \emph{exponentially small error} for min-entropy $k=Cdn^{C/d}$ for some constant \(C>0\), even when \(d\) is a constant.\footnote{Note that \cref{thm:lowdegsumsetext} can handle min-entropy \(k=O(dn^{1/\lfloor d/2\rfloor})\) if \(d\geq O(\log(1/\eps))\).} We think that even showing this for small linear min-entropy would already be quite interesting.

    \item In \cref{thm:lowErrorLowDeg2Ext}, we showed that there exist degree $\leq 4$ low-error two-source extractors for any linear min-entropy via approximate duality.
    This approach, however, provably cannot go below min-entropy $\sqrt{n}$~\cite{BLR12}.
    Can we show the existence of low-degree low-error two-source extractors for min-entropy below $\sqrt{n}$?
\end{itemize}

\dobib

\section*{Acknowledgements}

We thank Alexander Golovnev, Zeyu Guo, Pooya Hatami, Satyajeet Nagargoje, and Chao Yan for sharing with us an early draft of their work. We also thank Dean Doron for insightful discussions and feedback, and Mohit Gurumukhani for helpful pointers to facts about quadratic forms. Part of this work was carried out while the authors were visiting the Simons Institute for the Theory of Computing at UC Berkeley, and while the second and fourth author were visiting NTT Research in Sunnyvale, CA.

\newpage

\bibliographystyle{alpha}
\bibliography{./references}

\newpage
\appendix

\section{Existential results for sumset extractors via a uniformly random function}\label{sec:randsumsetext}

As a special case of our second main result (\cref{thm:lowdegsumsetext}), we know that a random degree \(d=O(\log(n/\eps))\) polynomial is an \(\eps\)-extractor for \((n,k)\)-sumset sources with min-entropy \(k=O(\log(n/\eps))\). A natural question is whether there is a simpler proof that a uniformly random function can achieve this min-entropy requirement. In this section, we show that the answer is yes, and give a simple proof of the following.

\begin{restatable}[Existential result for sumset extractors, originally proved in~\cite{Mra16}]{theorem}{existsumsetext}\label{thm:existsumsetext}
For every constant \(\gamma>0\) there exists a constant \(C>0\) such that for any \(n\geq k\in\N\) and \(\eps>0\) such that \(k\geq(2+\gamma)\log n + (4+2\gamma)\log(1/\eps)+C\), the following holds. A uniformly random function \(f:\F_2^n\to\F_2\) is a \((k,\eps)\)-sumset extractor with probability at least \(0.99\).
\end{restatable}

\cref{thm:existsumsetext} was posed as an open problem by Chattopadhyay and Liao at STOC 2022~\cite{CL22}, and the previous version of our work claimed to resolve it for the first time.\footnote{In fact, in the previous version of our work, \cref{thm:lowdegsumsetext} only worked for polynomially small error, and thus did not show that degree \(d=O(\log(n/\eps))\) polynomials can extract from sumset sources with min-entropy \(k=O(\log(n/\eps))\) (as it does now). Thus, in the previous version of our work, \cref{thm:existsumsetext} was presented as a completely separate result that established the existence of sumset extractors for min-entropy \(k=O(\log(n/\eps))\), instead of as a ``simpler version'' of \cref{thm:lowdegsumsetext}.} However, it turns out that this result was already known in the Cayley sum graph literature since 2015.
Namely, Mrazovi\'c~\cite{Mra16} showed that a uniformly random function is, with high probability, a $(k,\eps)$-sumset extractor for $k=2\log n + 6\log(1/\eps)+O(1)$.
Later work by Konyagin and Shkredov~\cite{KS18} improved the constant in front of the $\log n$ term at the expense of a worse dependence on $\eps$.
More recently, Conlon, Fox, Pham, and Yepremyan~\cite{CFPY24} extended this to more groups (among other results).

Thus, the existential result presented in \cref{thm:existsumsetext} is not novel, and should be attributed to Mrazovi\'{c}. However, we opted to keep this section in the current version of our work, since our proof seems somewhat novel,\footnote{In particular, our proof follows immediately from a seemingly novel structural result for sumset sources (\cref{lem:addinter}). This structural result can be thought of as a simpler version of our key structural result used in the low-degree setting (\cref{lem:convexCombFullRank}), and the proof of \cref{lem:addinter} can be viewed as a simpler variant of the randomized convex combination approach used in the proof of \cref{lem:convexCombFullRank}. Moreover, as we will soon discuss, the structural result in \cref{lem:addinter} implies the (``one-sided'' version of the) main structural result used in the work of Mrazovi\'{c} \cite{Mra16}.} and because we believe the presentation here may nevertheless be helpful to other people in the field. In what follows, we give an overview of our proof of \cref{thm:existsumsetext} vs.\ the proof of Mrazovi\'{c}, and conclude with a formal presentation of our proof.

\paragraph{Our approach vs.\ the original proof of Mrazovi\'c}
Unlike many pseudorandom objects, it is not easy to show that a uniformly random function \(f:\F_2^n\to\F_2\) is a good extractor for \((n,k)\)-sumset sources. This is because a naive counting argument shows that there are \(\approx\binom{2^n}{2^k}\binom{2^n}{2^k}\geq2^{2(n-k)2^k}\) such sources, but some of these sources only have \(2^k\) elements in their support (for example, if the sumset source is actually an affine \((n,k)\)-source). Thus, even if one had access to a perfect concentration bound, a naive application of the probabilistic method is destined to fail (since \(2^{-2^k}\cdot2^{2(n-k)2^k}\geq1\)).

In order to get around this issue, the key observation is as follows. While it seems difficult to dramatically improve the naive count of \((n,k)\)-sumset sources, it does seem like \emph{most} \((n,k)\)-sumset sources should have \(\gg2^k\) elements in their support. And if we restrict our attention to \((n,k)\)-sumset sources of this type, then there is some hope that the naive probabilistic argument (outlined above) will go through.

As it turns out, this is exactly the case. In particular, if we only consider the family \(\mathcal{Y}\) of \((n,k)\)-sumset sources \(\W=\X+\Y\) with low \emph{additive energy},\footnote{Having low additive energy can be thought of as a stronger (statistical) version of having support size \(\gg2^k\) - see \cref{def:additive-energy-existential}.} then it is easy to show that a uniformly random function is an excellent extractor for \(\mathcal{Y}\). This idea first appeared in the work of Mrazovi\'c \cite{Mra16}, and subsequently in the work of Chattopadhyay and Liao \cite{CL22}.

The final (and main) step is to show (via a structural result) that an extractor for sumset sources with low additive energy is automatically an extractor for \emph{all} sumset sources. This is the step that is missing in \cite{CL22}, and provided by both Mrazovi\'c \cite[Proposition 4]{Mra16} and, subsequently, our work (\cref{lem:addinter}, below). We outline and compare these structural results, below.

For his structural result, Mrazovi\'c shows that for any fixed function \(f:\F_2^n\to\F_2\) and sets \(X,Y\subseteq\F_2^n\), there exist slightly smaller sets \(X^\star,Y^\star\subseteq\F_2^n\) that (1) have low additive energy, and (2) have the property that \(\mathsf{bias}(f(X+Y))\leq \mathsf{bias}(f(X^\star+Y^\star))+\gamma\) for some small \(\gamma\).\footnote{Mrazovi\'c actually shows the slightly stronger result \(|\mathsf{bias}(f(X+Y))-\mathsf{bias}(f(X^\star+Y^\star))|\leq\gamma\), but only the ``one-sided'' version, stated above, is actually needed/used.} As a result, it follows that any extractor for sumset sources with low additive energy automatically works for all sumset sources.\footnote{This can be seen by taking \(f\) in property (2), above, to be the extractor.} Mrazovi\'c shows the existence of such sets \(X^\star,Y^\star\) by taking them to be uniformly random subsets of \(X,Y\) (respectively) of an appropriate size.

For our structural result, we show that every sumset source \(\X+\Y\) is a convex combination of sumset sources \(\X^\star+\Y^\star\) with low additive energy. This also shows that any extractor for sumset sources with low additive energy automatically works for all sumset sources. Furthermore, it implies the (one-sided) structural result of Mrazovi\'c (even with \(\gamma=0\)), by an averaging argument over the participants in the convex combination. In order to obtain our structural result, we assume (without loss of generality) that \(\X,\Y\) are uniform over some sets \(X,Y\), and randomly partition these sets into subsets \(\{X_i^\star\},\{Y_i^\star\}\) that each have the same size. The participants in our convex combination are then taken to be all random variables of the form \(\X_i^\star+\Y_j^\star\), where \(\X_i^\star,\Y_j^\star\) are uniform over \(X_i^\star,Y_j^\star\).

\subsubsection*{A formal presentation of our approach}

We now present our formal proof of \cref{thm:existsumsetext}. For this, we need the notion of \emph{additive energy} between two sets \(X,Y\) (which is closely related to the collision probability, or R\'{e}nyi entropy \(H_2\), of their sum).
\begin{definition}[\protect{\cite[Definition 2.8]{tao2006additive}}]\label{def:additive-energy-existential}
    The \emph{additive energy} between two sets $X,Y\subseteq\F_2^n$ is defined as
    \begin{equation*}
        E(X,Y):=\sum_{w\in \F_2^n} |\{(x,y)\in X\times Y: x+y=w\}|^2 = |\{(x,y,x',y')\in X\times Y\times X\times Y:x+y=x'+y'\}|.
    \end{equation*}
\end{definition}

\cref{thm:existsumsetext} is an easy consequence of the following structural lemma, which states that every pair of independent flat sources can be exactly partitioned into pairs of independent flat sources with low additive energy with slightly lower min-entropy.
We prove this result by analyzing random partitions of the supports of the two independent sources.

\begin{lemma}\label{lem:addinter}
    Let $X,Y\subseteq\F_2^n$ be sets of size at least $2^k$ each.
    Then, we can partition $X$ and $Y$ into $T=2^t$ subsets $(X_i)_{i\in[T]}$ and $(Y_j)_{j\in[T]}$, respectively, such that $|X_i|,|Y_j|= 2^{k-t}$ and $E(X_i,Y_j)\leq \ell^2\cdot 2^{2(k-t)}$ for all pairs $(i,j)\in[T]^2$, provided that $\ell\geq 4$ and $t\geq \frac{k}{\ell-1}(1+\ell/2)$.

    In particular, this means that we can write $\X$ and $\Y$ uniformly distributed over $X$ and $Y$, respectively, as convex combinations $\X=\sum_{i}p_i\X_i$ and $\Y=\sum_j q_j \Y_j$ for random variables $\X_i$ and $\Y_j$ uniformly distributed over sets $X_i$ and $Y_j$, respectively, satisfying the properties detailed above.
\end{lemma}
\begin{proof}
    Fix $\ell\geq 4$ and $t\geq\frac{k}{\ell-1}(1+\ell/2)$.
    Note that these constraints guarantee that $t\leq k$.
    Consider randomly partitioning each of $X$ and $Y$ (without replacement) into $T=2^t$ subsets $X_1,\dots,X_T$ and $Y_1,\dots,Y_T$, respectively, of size exactly $2^{k-t}$.
    Define $A_w=\{(x,y)\in X\times Y:x+y=w\}$ and $A^{i,j}_w=\{(x,y)\in X_i\times Y_j:x+y=w\}$.
    We show that we will sample with positive probability partitions $(X_i)$ and $(Y_j)$ such that $|A^{i,j}_w|\leq \ell$ for all $w\in\F_2^n$ and pairs $(i,j)$.
    In turn this implies that
    \begin{equation*}
        E(X_i,Y_j)=\sum_{w\in\F_2^n}|A_w^{i,j}|^2\leq \ell^2\cdot 2^{2(k-t)},
    \end{equation*}
    since the set $A_w^{i,j}$ is non-empty for at most $|X_i|\cdot |Y_j|=2^{2(k-t)}$ choices of $w$, as desired.
    
    Fix a vector $w\in\F_2^n$ and distinct pairs $(x^{(1)},y^{(1)}),\dots,(x^{(\ell)},y^{(\ell)})\in A_w$.
    Note that, since $x^{(i)}+y^{(i)}=x^{(j)}+y^{(j)}$, it follows that $x^{(1)},\dots,x^{(\ell)}$ are pairwise distinct, and the same holds for $y^{(1)},\dots,y^{(\ell)}$.
    Therefore, the probability that the pairs $(x^{(1)},y^{(1)}),\dots,(x^{(\ell)},y^{(\ell)})$ all end up in the same product set $X_i\times Y_j$ for some $(i,j)$ is at most\footnote{This uses the fact that $\frac{a-1}{b-1}\leq \frac{a}{b}$ whenever $a\leq b$.}
    \begin{equation*}
        \left(\frac{1}{T^2}\right)^{\ell-1}=2^{-2t(\ell-1)}.
    \end{equation*}
    
    For each vector $w\in\F_2^n$, define $s_w=|A_w|$, and, for pairs $(x^{(a)},y^{(a)})_{a\in[\ell]}\in A_w^\ell$, let $\ZZ_{(x^{(a)},y^{(a)})_{a\in[\ell]}}$ be the indicator random variable of the event that $(x^{(a)},y^{(a)})\in X_i\times Y_j$ for all $a\in[\ell]$ and some indices $i,j\in[T]$.
    Then, taking the expectation over the sampling of the partitions $X_1,\dots,X_T$ and $Y_1,\dots,Y_T$, we have
    \begin{align*}
        \E\left[\sum_{w\in\F_2^n} \sum_{(x^{(i)},y^{(i)})_{i\in[\ell]}\in A_w^\ell}\ZZ_{(x^{(i)},y^{(i)})_{i\in[\ell]}}\right]&\leq \sum_{w\in\F_2^n} \binom{s_w}{\ell} 2^{-2t(\ell-1)}\\
        & \leq \sum_{w\in\F_2^n:s_w\neq 0} \binom{2^k}{\ell} 2^{-2t(\ell-1)}\\
        &< 2^{2k}\cdot 2^{k\ell} \cdot 2^{-2t(\ell-1)}\\
        &\leq 1.
    \end{align*}
    The second inequality uses the fact that $s_w\leq 2^k$ for all $w\in\F_2^n$.
    The third inequality holds because $\binom{2^k}{\ell}<2^{k\ell}$ when $\ell\geq 4$ and because $\sum_{w\in\F_2^n}s_w\leq 2^{2k}$, meaning that there are at most $2^{2k}$ vectors $w$ such that $s_w\neq 0$.
    The last inequality follows from the choice $t\geq\frac{k}{\ell-1}(1+\ell/2)$.
    This implies that there must exist a partition of $X$ and $Y$ into $X_1,\dots,X_T$ and $Y_1,\dots,Y_T$, respectively, such that the associated sum inside the expectation is $0$, which means that $|A^{i,j}_w|\leq \ell$ for all $i,j\in [T]$, as desired.
\end{proof}

This lemma guarantees that to extract from any $k$-sumset source it suffices to extract from any $k$-sumset source $\W=\X+\Y$ such that $\X$ and $\Y$ are uniformly distributed over sets $X$ and $Y$, respectively, of slightly smaller size and such that the additive energy $E(X,Y)$ is appropriately small.
The proof of \cref{thm:existsumsetext} is then a consequence of \cref{lem:addinter}, the following lemma of Chattopadhyay and Liao~\cite{CL22} (which shows that random functions extract from sumset sources with low additive energy), and a union bound.

\begin{lemma}[\protect{\cite[Lemma B.2]{CL22}}]
\label{lem:randfunclowaddinter}
For any independent flat $k$-sources $\X,\Y\sim\F_2^n$, a uniformly random function $f:\F_2^n\to\F_2$ satisfies $f(\X+\Y)\approx_\eps \U_1$ with probability at least $1-2\cdot 2^{-\frac{2\eps^2 2^{4k}}{E(X,Y)}}$.
\end{lemma}
We are now ready to prove \cref{thm:existsumsetext}.
\begin{proof}[Proof of \cref{thm:existsumsetext}]
    Without loss of generality, fix a $k$-sumset $\X+\Y\sim\F_2^n$ with $\X$ and $\Y$ flat $k$-sources.
    Set $\ell = 1+k(1/2-\alpha)$ for an arbitrary constant $\alpha>0$ and $t\geq \frac{k}{\ell-1}(1+\ell/2)= (1+\beta)\frac{k}{2}+O(1)$.
    According to \cref{lem:addinter}, we can then write
    \begin{equation*}
        \X+\Y=\sum_{i,j\in[T]} p_{i,j}(\X_i+\Y_j),
    \end{equation*}
    where $(\X_i,\Y_j)$ are independent flat $k'$-sources with $k'=k-t=(1-\beta)\frac{k}{2}-O(1)$ and supports $X_i$ and $Y_j$ satisfying $E(X_i,Y_j)\leq \ell^2\cdot 2^{2(k-t)}=\ell^2 \cdot 2^{2k'}$,
    for $T=2^t$ and some $(p_{i,j})_{i,j\in[T]}$.
    Therefore, it suffices to show that a uniformly random function $f:\F_2^n\to\F_2$ satisfies $f(\X'+\Y')\approx_\eps \U_1$ simultaneously for all pairs $(\X',\Y')$ of independent flat $k'$-sources with $E(\X',\Y')\leq \ell^2 2^{2k'}$ with the desired probability.
    
    \cref{lem:randfunclowaddinter} guarantees that, for a fixed pair $(\X',\Y')$ as above, $f(\X'+\Y')\approx_\eps \U_1$ holds for a uniformly random function $f$ with probability at least
    \begin{equation*}
        1-2\cdot 2^{-\frac{2\eps^2 2^{4k'}}{E(\X',\Y')}}\geq 1-2\cdot 2^{-\frac{2\eps^2 2^{2k'}}{\ell^2}}.
    \end{equation*}
    Since there are at most $\binom{2^n}{2^{k'}}^2\leq 2^{2n 2^{k'}}$ such pairs of sources, a union bound over all these pairs shows that a uniformly random function satisfies the desired property except with probability at most
    \begin{equation}\label{eq:probfailrandfuncsumset}
        2^{2n 2^{k'}}\cdot 2\cdot 2^{-\frac{2\eps^2 2^{2k'}}{\ell^2}} = 2\cdot 2^{2^{k'}\left(2n-\frac{2\eps^2 2^{k'}}{\ell^2}\right)}.
    \end{equation}
    Setting $k=(2+\gamma)(\log n + 2\log(1/\eps))+C$ for $\gamma>0$ an arbitrarily small constant and $C>0$ a sufficiently large constant gives that the term in \cref{eq:probfailrandfuncsumset} is strictly smaller than $1$. 
\end{proof}
 
\dobib

\end{document}